\documentclass{jfm}
\usepackage{graphicx}
\usepackage{epstopdf, epsfig}
\usepackage{subfigure}
\usepackage[usenames,dvipsnames]{xcolor}
\usepackage{mathptmx}
\usepackage{graphicx} 
\usepackage{fancyhdr}
\usepackage{fnpos}
\usepackage[english]{babel}
\usepackage{array}
\usepackage{droidsans}
\usepackage{charter}
\usepackage[T1]{fontenc}
\usepackage[usenames,dvipsnames]{xcolor}
\usepackage{setspace}
\usepackage[compact]{titlesec}
\usepackage{hyperref}
\usepackage{amssymb}
\usepackage{lipsum}

\shorttitle{Dip coating of bidisperse suspensions}
\shortauthor{D.-H. Jeong, M. Ka Ho Lee, V. Thi\'evenaz,
  M. Z. Bazant and A. Sauret}

\title{Dip-coating of bidisperse particulate suspensions}

\author{Deok-Hoon Jeong\aff{1},
  Michael Ka Ho Lee\aff{1},
  Virgile Thi\'evenaz\aff{1},
  Martin Z. Bazant\aff{2,3,4},
 \and Alban Sauret\aff{1}\corresp{\email{asauret@ucsb.edu}}}

\affiliation{\aff{1}Department of Mechanical Engineering, University of California, Santa Barbara, CA, USA
\aff{2}Department of Chemical Engineering, Massachusetts Institute of Technology, Cambridge, MA 02139, USA
\aff{3}Department of Mathematics, Massachusetts Institute of Technology, Cambridge, MA 02139, USA
\aff{4}Saint-Gobain Research North America, Northborough, MA 01532, USA}

\begin{document}

\maketitle

\begin{abstract}
Dip-coating consists in withdrawing a substrate from a bath to coat it with a thin liquid layer. This process is well-understood for homogeneous fluids, but heterogeneities such as particles dispersed in the liquid lead to more complex situations. Indeed, particles introduce a new length scale, their size, in addition to the thickness of the coating film. Recent studies have shown that at first order, the thickness of the coating film for monodisperse particles can be captured by an effective capillary number based on the viscosity of the suspension, providing that the film is thicker than the particle diameter. However, suspensions involved in most practical applications are polydisperse, characterized by a wide range of particle sizes, introducing additional length scales. In this study, we investigate the dip coating of suspensions having a bimodal size distribution of particles. We show that the effective viscosity approach is still valid in the regime where the coating film is thicker than the diameter of the largest particles, although bidisperse suspensions are less viscous than monodisperse suspensions of the same solid fraction. We also characterize the intermediate regime that consists of a heterogeneous coating layer and where the composition of the film is different from the composition of the bath. A model to predict the probability of entraining the particles in the liquid film depending on their sizes is proposed and captures our measurements. In this regime, corresponding to a specific range of withdrawal velocities, capillarity filters the large particles out of the film.
\end{abstract}


\section{Introduction}

Dip-coating is a common industrial coating method that consists in withdrawing a substrate from a liquid bath at a constant speed \cite[][]{scriven1988physics,ruschak1985coating,quere1999fluid,grosso2011exploit}. This method has been studied since 1942 by  \cite{levich1942dragging} and  \cite{derjaguin1943} in the configuration of a plate withdrawn at a constant velocity $U$ from a Newtonian liquid of viscosity $\eta$, density $\rho$, and surface tension $\gamma$. Far from the liquid bath, the thickness $h$ of the liquid film coating the plate is uniform and set by the balance of viscous stresses, which enable the plate to pull the liquid out of the bath, and capillary stresses at the meniscus, which pull the fluid back to the bath \cite[][]{rio2017withdrawing}. The relative magnitude of viscous stresses to capillary stresses at the meniscus is measured by the capillary number, ${\rm Ca}=\eta\,U/\gamma$. In the limit of small capillary number ${\rm Ca} \ll 1$ and small Reynolds number ${\rm Re} =\rho\,U\,h/\eta \ll 1$, the thickness of the coating film is given by the Landau-Levich-Derjaguin law:
\begin{equation} \label{eq:LLD}
h=0.94\,\ell_c\,{\rm Ca}^{2/3},
\end{equation}
where $\ell_c=\sqrt{\gamma/(\rho\,g)}$ is the capillary length. At larger capillary numbers, typically of the order of $Ca \gtrsim 10^{-2}$, gravity dominates capillary forces \cite[][]{maleki2011landau}. The balance between viscosity and gravity leads to a new scaling law for the thickness of the liquid film coating a plate, $h \propto \ell_c\,{\rm Ca}^{1/2}$.

Owing to the complexity of the fluids used in industrial processes, various studies have considered the dip-coating process for homogeneous fluids with complex rheology,
such as shear-thinning fluids \cite[][]{gutfinger1965films,hewson2009model},
yield-stress fluids \cite[][]{maillard2014solid,maillard2015flow,smit2019stress},
viscoelastic fluids \cite[][]{ro1995viscoelastic,de1998fluid,ruckenstein2002scaling},
as well as the influence of surfactants 
\cite[][]{shen2002fiber,krechetnikov2006surfactant,delacotte2012plate},
roughness \cite[][]{krechetnikov2005experimental,seiwert2011coating},
and the geometry of the substrate \cite[][]{white1965static,zhang2021dip}. Nevertheless, most of these studies consider homogeneous fluids, made of a single phase.
Suspensions, in which solid particles are dispersed in a liquid phase, are of particular interest to manufacturing applications. The particles can give specific properties to a surface after coating. Thus, dip-coating, in particular combined with evaporation, has been considered for optical applications, self-assembling of particles, and wettability treatments \cite[][]{ghosh2007spontaneous,mechiakh2010correlation, berteloot2013dip,mahadik2013superhydrophobic}.
More recently, several studies have considered the dip-coating of monodisperse suspensions (single particle size), of non-Brownian particles (diameter $d$ larger than a few tens of microns), in non-volatile liquids \cite[][]{kao2012spinodal,gans2019dip,palma2019dip}.
These studies revealed that depending on the withdrawal velocity $U$, the fluid properties, and the size of the particles, three different coating regimes are observed: 
(i) at small withdrawal velocity, a thin film is deposited without any particles in it;
(ii) at large withdrawal velocity (\textit{i.e.}, large capillary numbers), the entrained film contains particles, and its thickness follows the Landau-Levich law using at first order the effective viscosity of the suspension; 
finally (iii) at intermediate withdrawal velocities, the coating is heterogeneous, with an average film thickness that corresponds to a monolayer of particles and remains roughly constant over a range of capillary numbers.
For a monodisperse suspension, the transition between the different regimes is governed by the thickness of the coating film relative to the particle diameter $h/d$ \cite[][]{gans2019dip,palma2019dip}.

The transition between the no-particle and heterogeneous coating regime also depends on the accumulation of particles at the meniscus. This transition is therefore complex to predict quantitatively for non-dilute suspensions, typically as soon as the volume fraction, defined as $\phi=V_{\rm p}/(V_{\rm p}+V_{\rm l})$, where $V_{\rm p}$ and $V_{\rm l}$ are the volume of particles and liquid, respectively, is larger than a few percent.
The configuration of isolated particles is simpler to describe because the particles do not interact with each other. This configuration was considered for flat plates \cite[][]{colosqui2013hydrodynamically,sauret2019capillary} and fibers \cite[][]{dincau2020entrainment}.
These studies have shown that an isolated particle of diameter $d$ can be entrained in a coating film if the particle radius is roughly smaller than the thickness at the stagnation point $h^*$. Indeed, the stagnation point defines the boundary between a shear flow where the fluid continues into the coating film, and a recirculation flow where the fluid returns into the liquid bath.
Thus, the thickness at the stagnation point $h^*$ controls the entrainment of particles in the coating film \cite[][]{colosqui2013hydrodynamically}. The value of $h^*$ is related to the thickness of the coating film through  ${h^{*}}/{\ell_{c}}=3\,{h}/{\ell_{c}}-\left({h}/{\ell_{c}}\right)^{3}$  \cite[][]{levich1942dragging}, which in the limit of small capillary numbers becomes $h^{*}=3\,h$.
Experimentally, isolated particles of diameter $d$ are entrained in the coating film if \cite[][]{sauret2019capillary,dincau2020entrainment}:

\begin{equation}
\label{eq:condition_entrainment} h^{*}=3\,h\gtrsim d/2.
\end{equation}

The ability to control the film thickness, and thus the thickness at the stagnation point, by simply tuning the withdrawal velocity $U$ has led to a method for sorting particles by size through dip-coating \cite[][]{dincau2019capillary}. This study has considered dilute suspension and has shown that, since smaller particles can be entrained for smaller coating thickness, isolated particles can be separated by size via selecting an appropriate withdrawal velocity. 

This entrainment process is not specific to dip-coating. The translation of an air bubble in a tube, as well as the withdrawal of the fluid leading to the deposition of a thin film on the wall of a capillary tube, share many common features with the dip-coating configuration, in particular, the presence of a stagnation point  \cite[][]{bretherton1961motion,krechetnikov2010application}.
Therefore, similar observations on the entrainment of particles \cite[][]{jeong2020deposition,wu2021film} and the filtering of particles \cite[][]{yu2018separation} have been reported. 
We should also emphasize that the influence of particles on different interfacial phenomena, such as the formation of droplets \cite[][]{furbank2004experimental,bonnoit2012accelerated,chateau2018pinch,thievenaz2021pinch}, jets \cite[][]{chateau2019breakup}, and liquid sheets \cite[][]{raux2020spreading}, have also reported that the critical length scale at which the particles start to modify 
significantly the dynamics is comparable to the diameter of the particles.

Whereas most of these studies have considered the ideal situation of a suspension made of monodisperse particles, many industrial and environmental processes involve polydisperse particles with a wide range of sizes. It is known that for a given solid volume fraction, a polydisperse suspension will be less viscous
than its monodisperse counterpart \cite[][]{shapiro1992random}.
For dip-coating, the size distribution of the particles also needs to be compared to the thickness of the coating film. It remains unclear how the three regimes reported previously for monodisperse suspensions will need to be modified to account for the polydispersity of the suspension.

\begin{figure}
\centering
  \includegraphics[width=\textwidth]{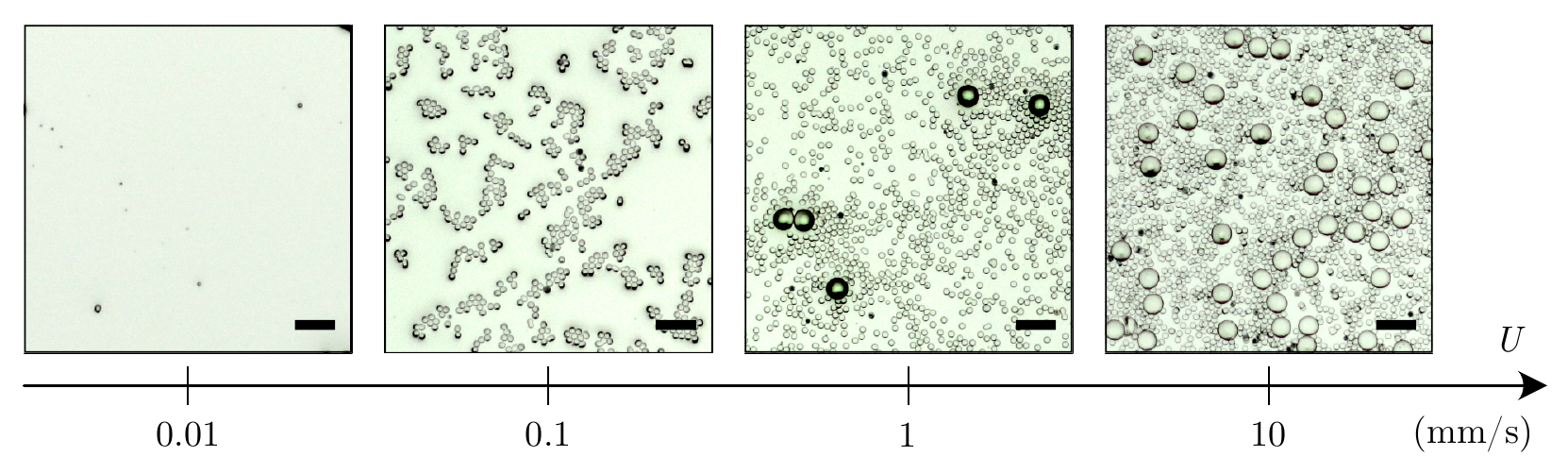}
  \caption{Typical coating films observed on a flat plate for increasing withdrawal velocities
    for a bidisperse suspension of particles of diameters $d_{\rm S}=\,250\,\mu{\rm m}$ and 
    $d_{\rm L}=\,80\,\mu{\rm m}$ (size ratio $\delta=3.125$), 
    at a volume fraction of $\phi=0.2$ and a volume ratio of large particles $\zeta=0.6$.
    The withdrawal velocity $U$ increases from left to right: 
    $U=0.01, 0.1, 1,$ and $10\,{\rm mm.s^{-1}}$.
    The size of the scale bars is 500$\,\mu\rm{m}$.}
  \label{fgr:Figure_1}
\end{figure}

Figure \ref{fgr:Figure_1} shows four examples of coating films on a plate withdrawn from a bidisperse suspension when increasing the withdrawal velocities $U$. The suspension contains particles of diameter $d_{\rm S}=250\,\mu{\rm m}$ and $d_{\rm L}=80\,\mu{\rm m}$, at a volume fraction of $\phi=0.2$. The volume ratio of large particles is $\zeta=V_{\rm L}/(V_{\rm L}+V_{\rm S})=0.6$, where $V_{\rm L}$ and $V_{\rm S}$ are the volume of large and small particles in the suspension, respectively. The main features observed for monodisperse suspensions are also observed with bidisperse suspensions. In particular, at very low withdrawal velocity ($U=0.01\,{\rm mm.s^{-1}}$), the particles remain in the liquid bath as they are much larger than the coating film. As a result, the meniscus filters them out, and the thin film is only made of liquid. At large withdrawal velocities ($U=10\,{\rm mm.s^{-1}}$) we observe an effective viscosity regime. Both populations of particles are present, in proportions similar to the suspension in the bath. At intermediate withdrawal velocities, a behavior specific of bidisperse suspensions is observed. Initially, at low withdrawal velocity ($U=0.1\,{\rm mm.s^{-1}}$), only small particles are present in the coating film. By increasing the withdrawal velocity ($U=1\,{\rm mm.s^{-1}}$), the thickness of the coating film also increases, resulting in more and more large particles being entrained in the film. As a result, the composition of the coating film differs from that of the bath in this regime, with varying proportions of small and large particles depending on the withdrawal velocity.

In this study, we aim to describe the evolution of the thickness and the composition of the coating film when varying the capillary number and the composition of the suspension. As a first step towards polydisperse systems, we consider bidisperse suspensions made of small and large particles of diameter $d_{\rm S}$ and $d_{\rm L}$, respectively. The volume ratio of large to small particles is varied to probe the influence of the size distribution of particles on the formation and composition of the coating film. This paper is organized as follows: The experimental methods and the suspensions used are first presented in section \ref{sec:exp}.
The dip-coating with monodisperse suspensions is recalled in section \ref{sec:mono}, notably to refine the measurements of the thickness of the coating film. Indeed, in the effective viscosity regime, we show that the volume fraction in the film is slightly smaller than in the suspension bath. Section \ref{sec:bi} is devoted to the experimental characterization with bidisperse suspension. We describe and rationalize the different regimes observed and show that in the thick-film regime, rheological models developed for bidisperse suspensions enable us to model the thickness of the coating film, while the heterogeneous regime is more complex for bidisperse suspension. We show that the composition of the coating film evolves with the withdrawal velocity, and we propose a model that captures the evolution of the composition of the coating film, in particular, the filtration of large particles at intermediate velocities.


\section{Experimental methods} \label{sec:exp}

\begin{figure}
\centering
 \subfigure[]{\includegraphics[width=0.75\textwidth]{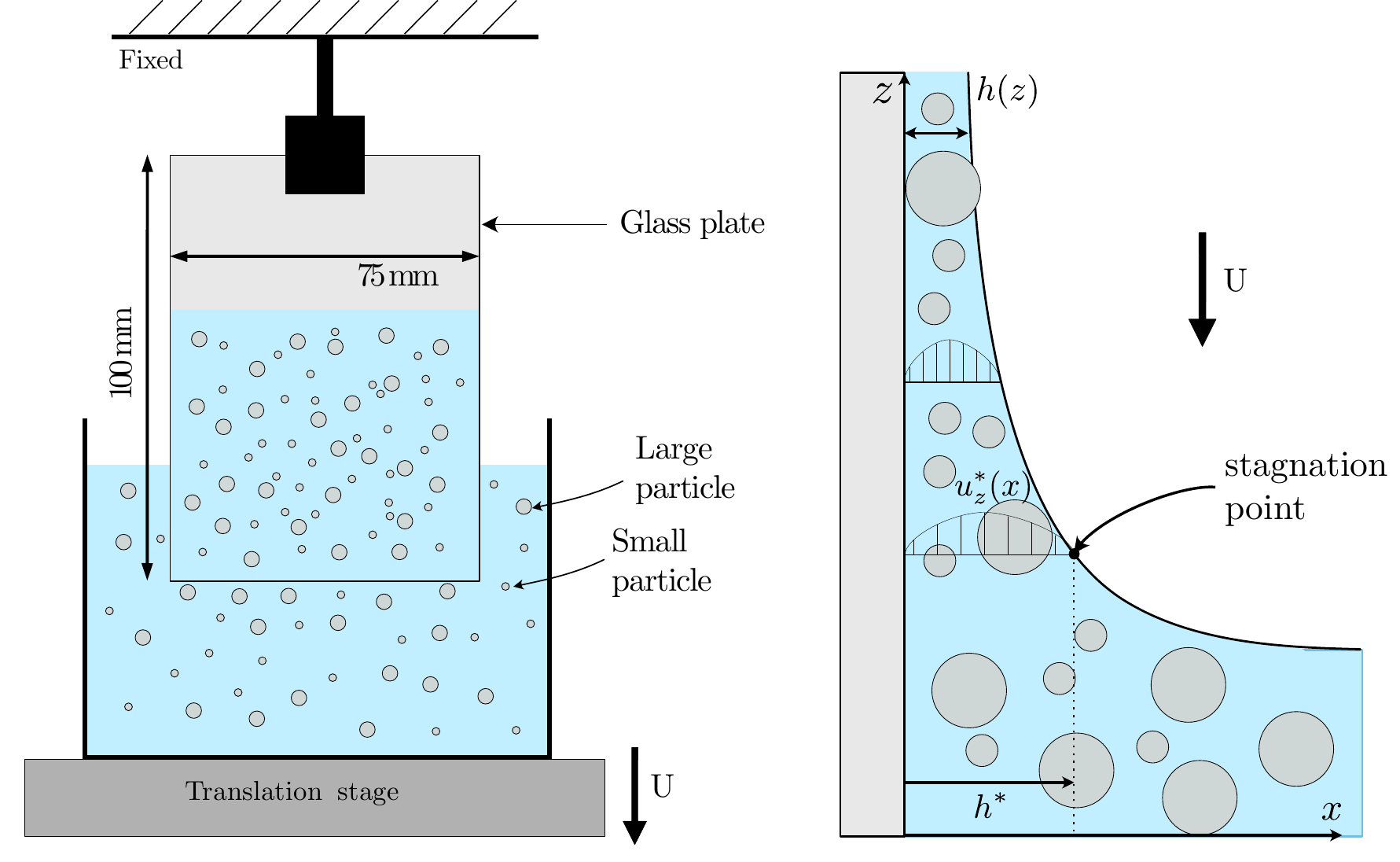}} \\
  \subfigure[]{\includegraphics[width=0.65\textwidth]{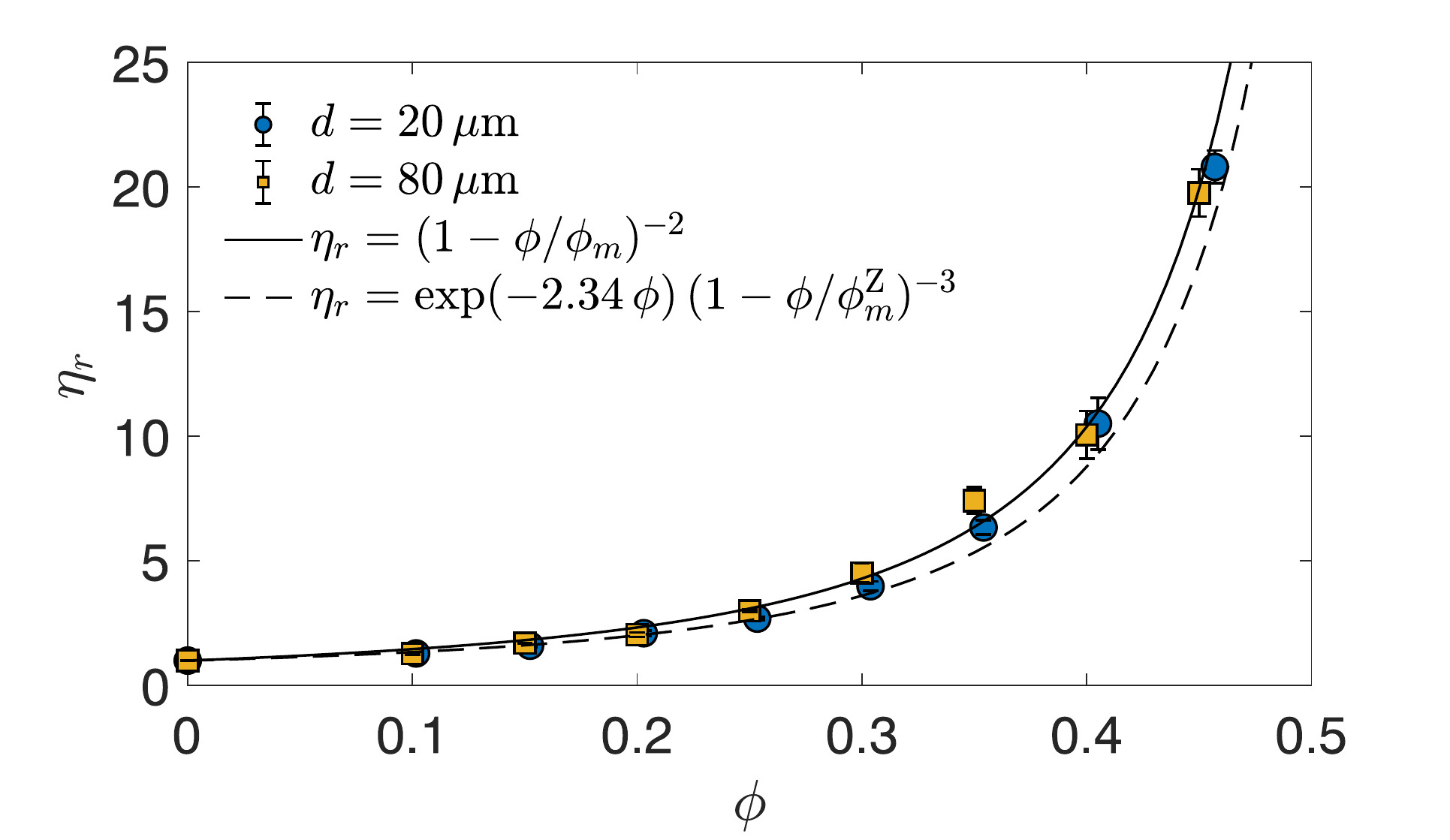}}
  \caption{(a) Schematic of the experimental setup.
      Front (left) and side view (right).
      (b) Relative effective shear viscosity $\eta_{\rm r}=\eta/\eta_0$ of monodisperse suspensions
      for particles of diameter $d=20\,\mu{\rm m}$ (blue circles) 
      and $d=80\,\mu{\rm m}$ (yellow squares).
      The solid line indicates the Maron–Pierce correlation [equation (\ref{eq:MP}) with $\phi_m=0.58$],
      and the dashed line is the Zarraga correlation [equation (\ref{eq:Z}) with $\phi_m^Z=0.62$].
  }
  \label{fgr:Figure_2}
\end{figure}

Our experiments consist in withdrawing a glass plate (75 mm wide and 3.25 mm thick) from a rectangular container (width 108 mm and thickness 35mm) filled with a particulate suspension. Figure \ref{fgr:Figure_2}(a) shows a schematic of the experimental setup. The suspensions are prepared by dispersing the non-Brownian particles in a silicone oil having a density close to the density of the particles. The particles used are spherical polystyrene particles (Dynoseeds TS from Microbeads) of diameter $d = 22,\, 81,\, 145$ and $249\,\mu\rm{m}$  (later referred as $d = 20,\, 80,\, 140$ and $250\,\mu\rm{m}$) and densities between $1046\, \rm{kg \cdot m^{-3}}$ and $1062\, \rm{kg\cdot m^{-3}}$ depending on the batch (see the physical characterization in Appendix). The silicone oil (AP100, Sigma-Aldrich) has a viscosity of $\eta_0 = 112 \rm{mPa \cdot s}$, density $\rho = \rm{1058 \,kg \cdot m^{-3}}$ and a surface tension of $\gamma=25 \pm 2\,{\rm mN \cdot m^{-1}}$ at $20^{\rm o}$C. Silicone oil perfectly wets the plate and the particles and is used for dip-coating experiments to avoid any potential effects from surfactants, which are known to increase the thickness of the coating film even at low concentrations \cite[][]{krechetnikov2005experimental,krechetnikov2006surfactant,rio2017withdrawing}. The particles are first dispersed using a paint mixer. Then, the suspension is left in a vacuum chamber for a few minutes to remove any entrapped bubble in the suspension. Between each experiment, the suspension is re-homogenized to ensure that the settling of the particles is negligible at the time scale of one experiment (typically a few minutes).

The liquid bath is placed on a stage that is translated vertically using a stepper motor (Thorlabs NRT150) at a given velocity $0.01\,{\rm mm.s^{-1}}<U<15\,{\rm mm.s^{-1}}$. Such an approach avoids mechanical perturbations that could influence the thickness of the coating film \cite[][]{maleki2011landau}. After the plate has been withdrawn from the liquid bath, pictures of the coating film are taken using a DSLR camera (Nikon D5600) equipped with a macro lens (Nikkor 200 mm). A microscopic lens (Mitutoyo M Plan Apo 5X) is also used for suspensions of $d = 20\,\mu{\rm m}$ particles. Between each experiment, the glass plate is thoroughly cleaned with isopropyl alcohol, rinsed multiple times with deionized water, and then dried with compressed air.

In addition to directly observing the coating film, its thickness is estimated by a gravimetric method, chosen for its excellent accuracy \cite[][]{krechetnikov2005experimental}. The liquid bath is placed on an analytical weighing scale (Ohaus SPX622 Scout, with an accuracy of $0.01\,{\rm g}$) during the experiments. The translating stage is moved up until the plate is dipped in the suspension bath by the desired dipping length $L_1$, and then withdrawn, holding a mass of entrained fluid $m_1$. The plate is then dipped again by a larger length $L_2$, and then withdrawn while holding an increased fluid mass $m_2$ on the plate. The resulting average thickness of the deposited liquid film is then given by

\begin{equation}
h= \frac{m_2-m_1}{\left(L_2-L_1\right) \rho\,\,P},
\end{equation}

\noindent where $P=2\,(w+e)$ is the perimeter of the plate, $w$ is the width and $e$ the thickness of the plate, and $\rho$ is the density of the suspension. Subscripts $1$ and $2$ denote the two dipping lengths such that $L_2>L_1$. This approach prevents a lower edge effect that interferes with the estimation of the film thickness \cite[][]{krechetnikov2005experimental}. For the width of the plate, the dipping lengths, and the scale used here, the uncertainty on the film thickness is of order $\pm 3\,\mu{\rm m}$. 
More details on this method have been provided by \cite{krechetnikov2005experimental}, and has previously been used with suspensions  \cite[][]{gans2019dip}.

The shear viscosity of the suspensions is measured using a dynamic shear rheometer (Anton Paar MCR92) with a 25 mm diameter plate-plate rough geometry and a gap of 1 mm between the plates. In the range of volume fraction considered here, the suspension has a Newtonian behavior and is characterized by its shear viscosity $\eta$. Figure \ref{fgr:Figure_2}(b) reports the evolution of the relative shear viscosity, $\eta_{\rm r}(\phi)=\eta(\phi)/\eta_0$ for a volume fraction in the range $10\%<\phi<40\%$ and two particles sizes ($20\,\mu{\rm m}$ and $80\,\mu{\rm m}$, monodisperse suspensions). Many empirical correlation between $\eta_{\rm r}$ and $\phi$ can be found in the literature \cite[][]{stickel2005fluid,guazzelli2018rheology}.
In the following, we use the Maron-Pierce correlation:

\begin{equation}\label{eq:MP}
    \eta_{\rm r}=\frac{\eta(\phi)}{\eta_0}=(1-\phi/\phi_{\rm m})^{-2},
\end{equation}

\noindent where $\phi_{\rm m}$ corresponds to the volume fraction of particles at which the viscosity diverges. Fitting equation~(\ref{eq:MP}) to our measurements leads to $\phi_{\rm m} \approx 0.58$, in agreement with other measurements performed in the literature with the same particles \cite[][]{guazzelli2018rheology,chateau2018pinch}. Note that another correlation can be used. For instance, the Zarraga correlation \cite[][]{zarraga2000characterization} has been used to describe the dip-coating of monodisperse suspensions \cite[][]{gans2019dip}, the pinch-off of suspension droplets \cite[][]{bonnoit2012accelerated}, and the flow of suspension on an inclined plane \cite[][]{bonnoit2010inclined}. The Zarraga correlation is given by

\begin{equation}\label{eq:Z}
\eta_r=\frac{\eta(\phi)}{\eta_0}=\frac{\exp (-2.34 \phi)}{\left(1-\phi / \phi_{\rm m}^{\rm Z}\right)^{3}},
\end{equation}

\noindent which leads with our measurements to $\phi_{\rm m}^{\rm Z}=0.62$, also in agreement with the values reported in other studies \cite[][]{bonnoit2012accelerated}. The Zarraga correlation slightly underestimates the effective shear viscosity for 
$\phi \gtrsim 0.25$ but better captures it at moderate volume fraction ($\phi \sim 0.1-0.2$). The Eilers correlation is also an option \cite[][]{stickel2005fluid}, and has been used recently for the spreading of suspension droplets \cite[][]{zhao2020spreading}. Our decision to choose the Maron-Pierce correlation here is motivated by previous studies showing that the viscosity diverges as $(1-\phi/\phi_{\rm c})^{-2}$, stressing the exponent $-2$ \cite[][]{guazzelli2018rheology}.

For a given $\phi$, the effective shear viscosity of a bidisperse suspension is lower than that of a monodisperse suspension \cite[][]{shapiro1992random,probstein1994bimodal,gamonpilas2016shear,guy2020testing}. This effect is linked to the higher compacity of polydisperse sphere packings \cite[][]{ouchiyama1984porosity}. Indeed, in a packing of polydisperse spheres, small particles can fill the interstices between the larger ones, which leads to a higher maximum packing fraction $\phi_{\rm m}$. Compared to the monodisperse case where $\eta$ is only a function of $\phi$,
the viscosity of bidisperse suspensions depends on two additional parameters:
the ratio of large to small particle diameters $\delta = d_{\rm L}/d_{\rm S}$, and the fraction of the solid volume occupied by the larger particles $\zeta = V_{\rm L}/(V_{\rm L} + V_{\rm S})$ \cite[][]{shapiro1992random}. Experimental measurements have shown that the viscosity of bidisperse suspensions follows the Maron-Pierce correlation, provided that $\phi_{\rm m}$ takes the polydispersity into account \cite[][]{thievenaz2021droplet}.


\section{Dip-coating of monodisperse suspension} \label{sec:mono}

Monodisperse suspensions, \textit{i.e.}, composed of particles of a single size, are first considered for volume fractions ranging from $\phi=10\%$ to $\phi=40\%$ and different particle diameters. The goal here is to verify whether the Maron-Pierce correlation [equation (\ref{eq:MP})] can predict the thickness of the coating films. \cite{gans2019dip} and \cite{palma2019dip} have previously shown that if the film is approximately thicker than the particle diameter ($h \gtrsim d$), its thickness follows the same law as a viscous liquid [equation~(\ref{eq:LLD})], where the viscosity corresponds to the effective viscosity of the suspension. However, despite a good agreement, this approach slightly overestimates the thickness of the coating film (see figure 5 and figure 8 for large volume fraction in \cite{gans2019dip}).

\begin{figure}
    \centering
    \subfigure[]{\includegraphics[width=0.495\textwidth]{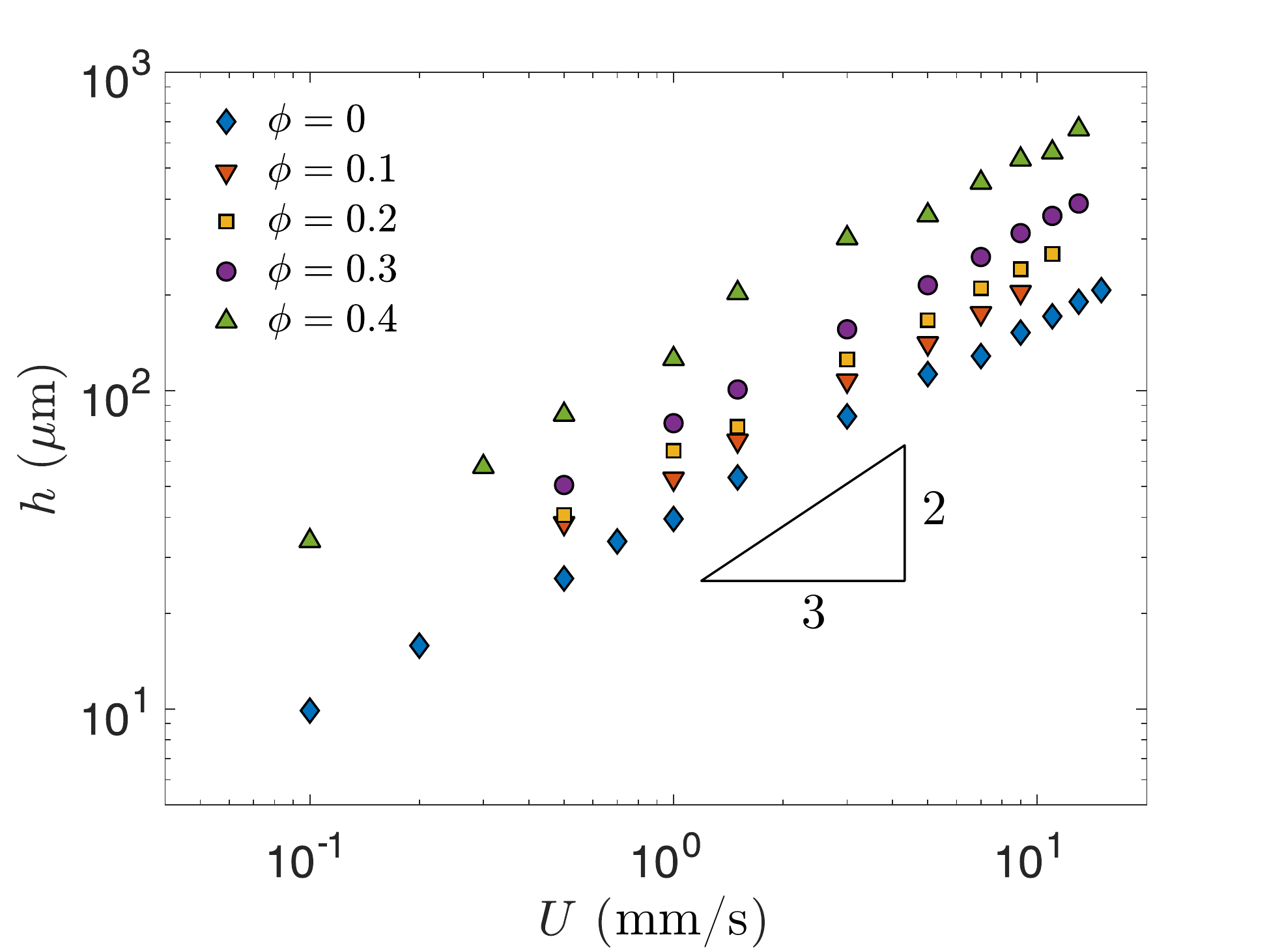}}
    \subfigure[]{\includegraphics[width=0.495\textwidth]{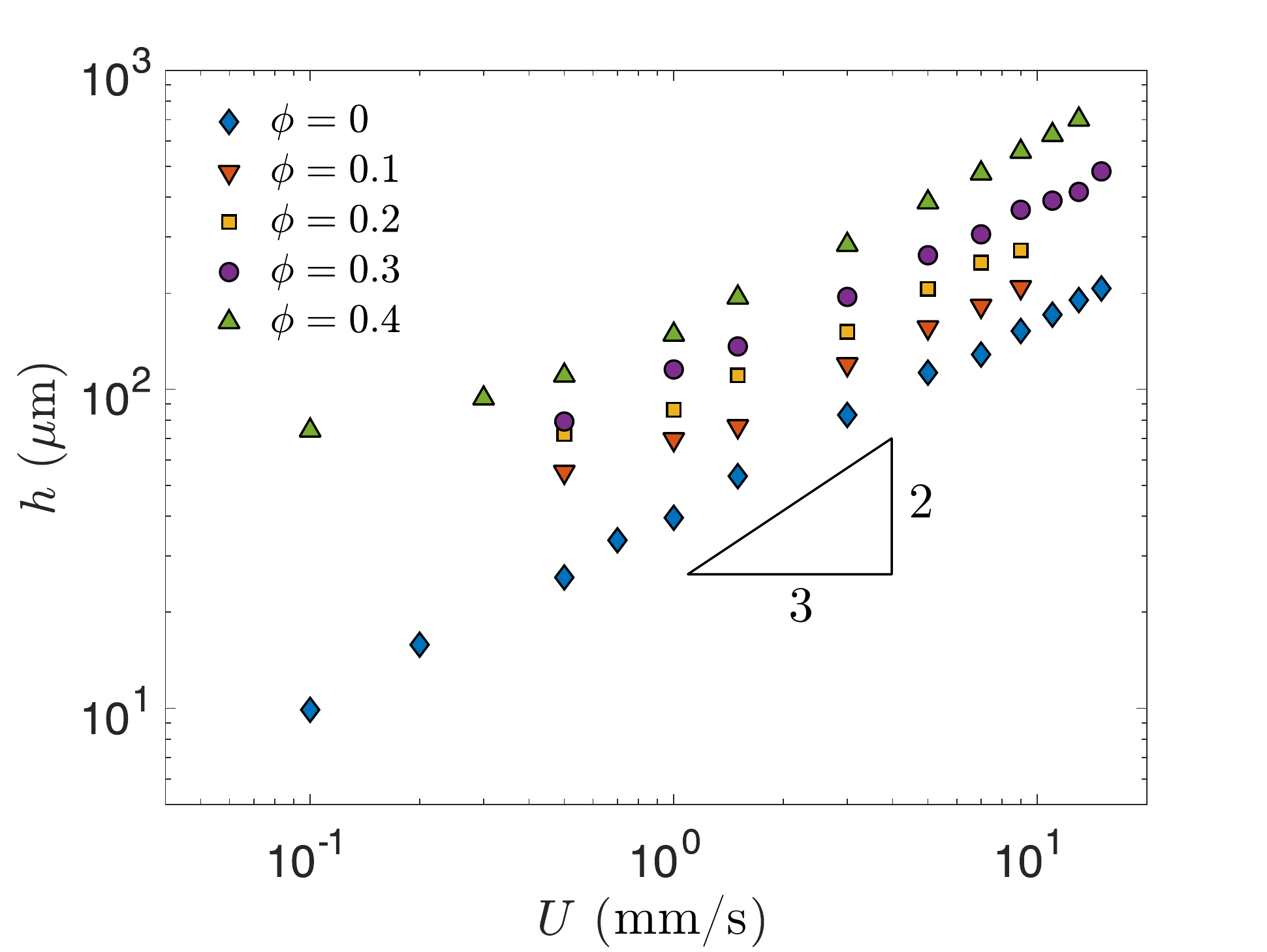}}
    \subfigure[]{\includegraphics[width=0.495\textwidth]{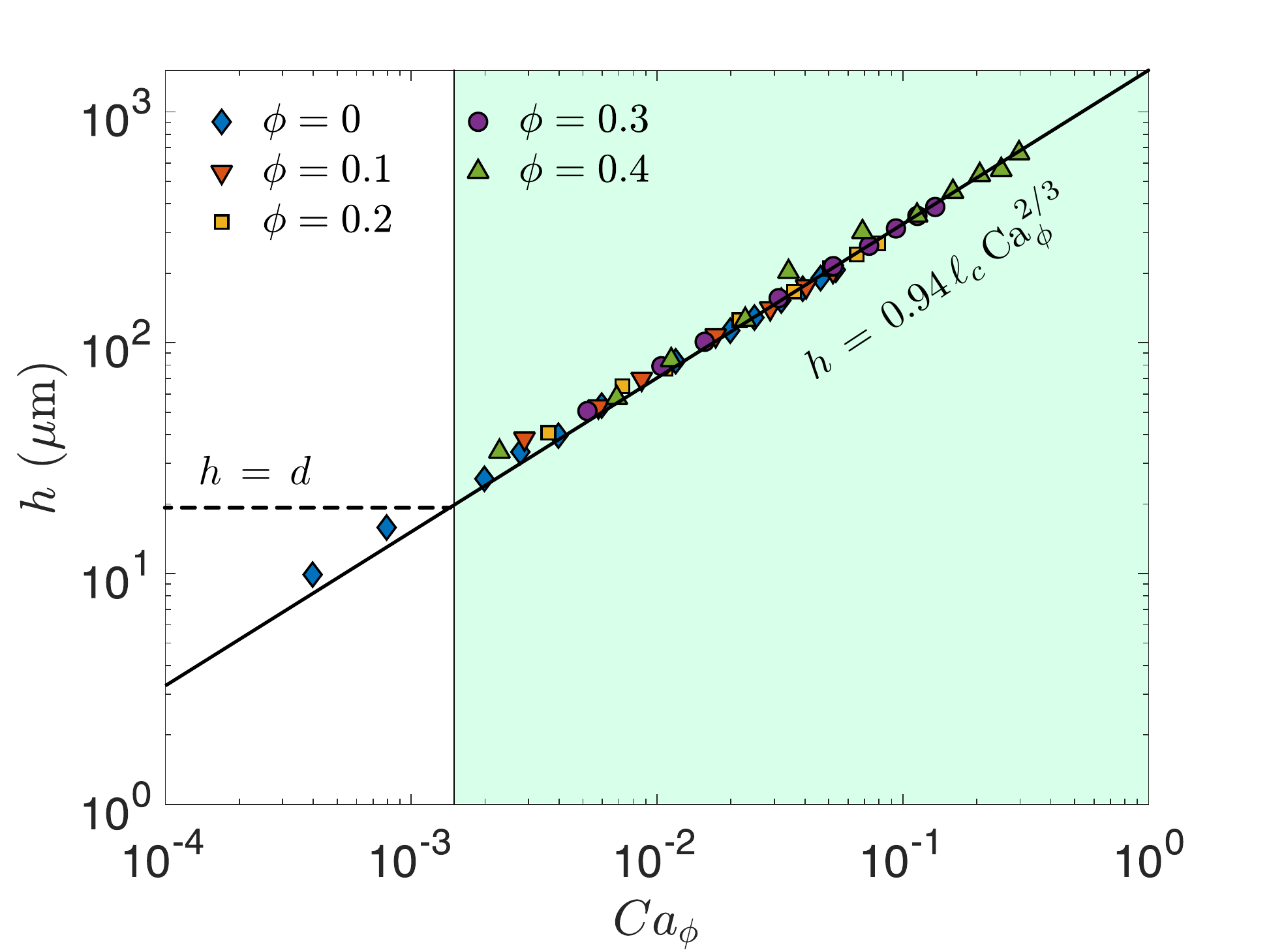}}
    \subfigure[]{\includegraphics[width=0.495\textwidth]{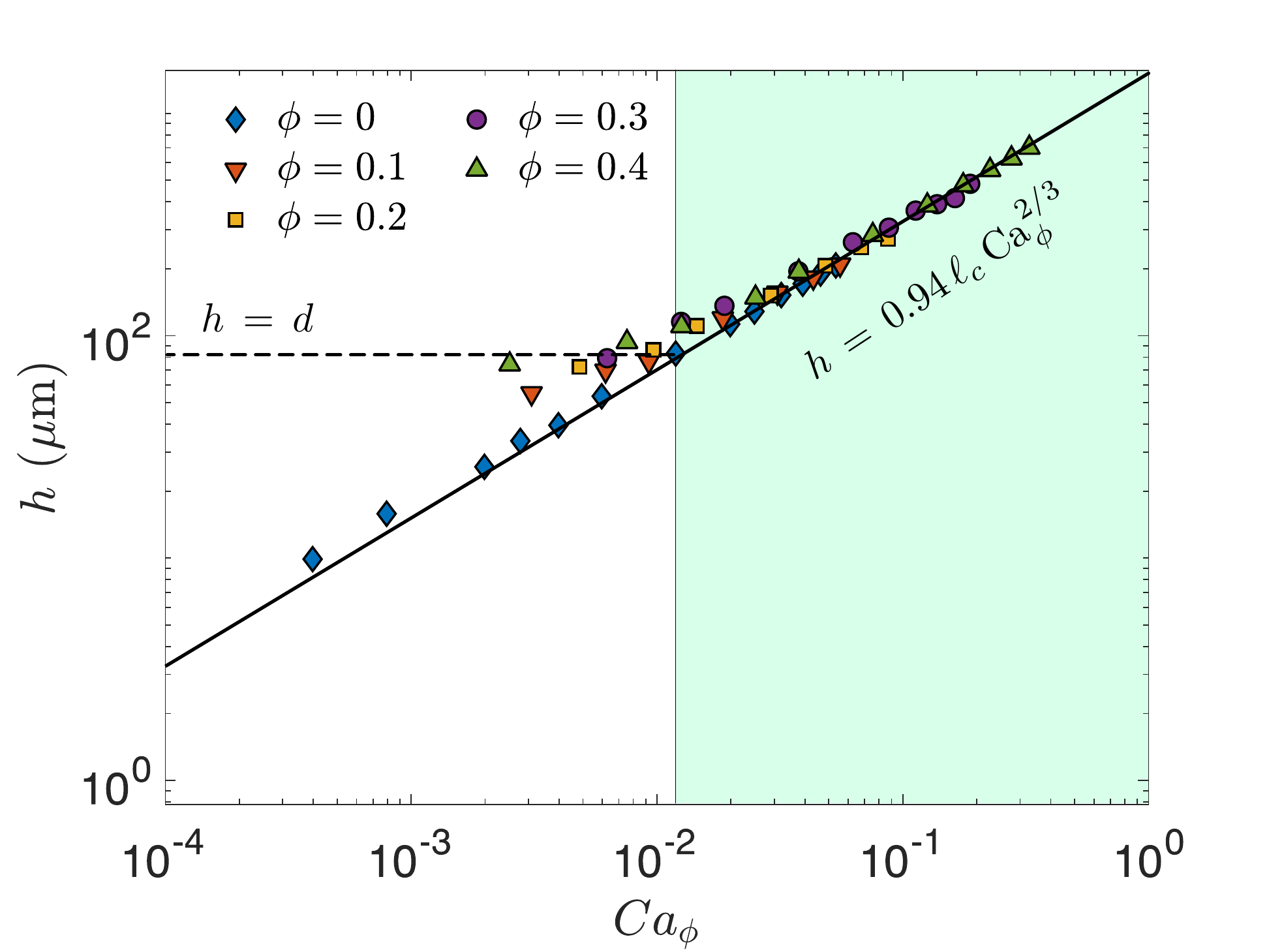}}
    \caption{Thickness of the coating film as a function of (a)-(b) the withdrawal velocity $U$, and of
        (c)-(d) the effective capillary number ${\rm Ca}_{\phi}$, for varying volume fractions $\phi$ of particles of diameter (a)-(c) $d = 20\,\mu{\rm m}$, and (b)-(d) $d = 80\,\mu{\rm m}$. The thick continuous line is the LLD law [equation \ref{eq:LLD}] where the viscosity is considered as a fitting parameter. The horizontal dashed line in figures (c) and (d) corresponds to a coating film as thickness equals to the particle diameter ($h=d$). The colored area in (c) and (d) corresponds to the effective viscosity regime.}
    \label{fgr:Figure_3}
\end{figure}

Figures \ref{fgr:Figure_3}(a) and \ref{fgr:Figure_3}(b) show the thickness of the coating film $h$ when varying the withdrawal velocity of the plate $U$ for particles of diameter $20\,\mu{\rm m}$ and $80\,\mu{\rm m}$, respectively. As expected, the faster the withdrawal, the thicker the coating film. Besides, increasing the volume fraction of particles, and thus the viscosity of the suspension in the bath, also leads to thicker films. When $h \gtrsim d$, we observe the transition to the effective viscosity regime, in which $h \propto U^{2/3}$ according to the LLD law [equation~(\ref{eq:LLD})], in agreement with previous works \cite[][]{gans2019dip,palma2019dip}.

To begin with, no assumption regarding the effective viscosity of the suspension is made. Instead, it is treated as a fitting parameter that we can estimate through the Landau-Levich-Derjaguin law:

\begin{equation}
    h=0.94\,\ell_c\,{{\rm Ca}_\phi}^{2/3}=0.94\,\ell_c\,\left(\frac{\eta(\phi)\,U}{\gamma}\right)^{2/3},
    \label{eq:LLD_effVisc}
\end{equation}

The capillary number is based on the effective viscosity of the suspension: ${\rm Ca}_\phi=\eta(\phi)\,U/\gamma$. In this expression, the capillary length $\ell_c$ and the surface tension $\gamma$ are physical properties of the liquid which are not modified by the particles, $U$ is the withdrawal velocity, and $\eta(\phi)$ is the effective viscosity of the suspension. For each experiment, we compute the value of $\eta(\phi)$ so that the thickness of the film in the LLD regime is captured by equation~(\ref{eq:LLD_effVisc}).

\begin{figure}
    \centering
    \subfigure[]{\includegraphics[width=0.495\textwidth]{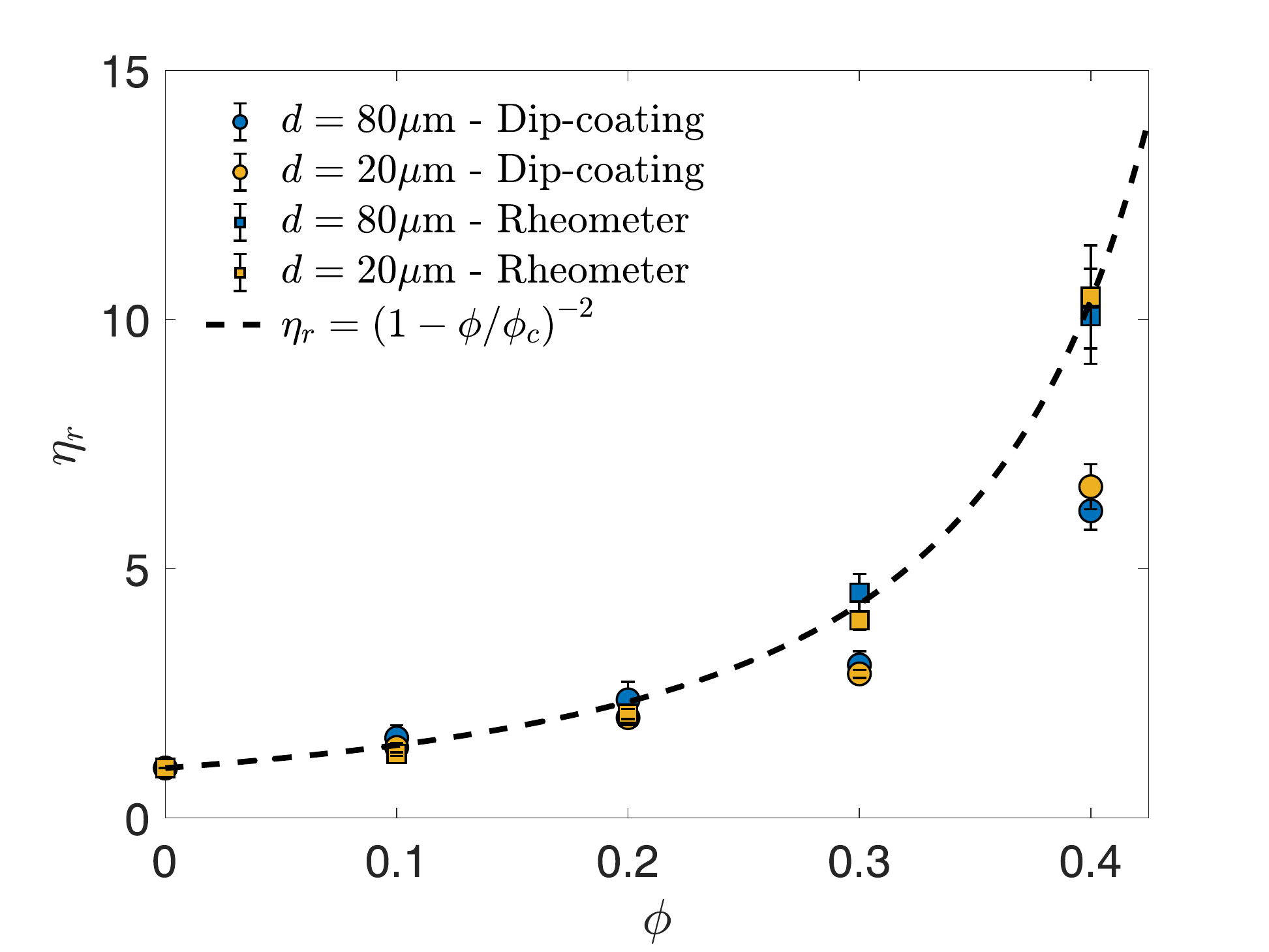}}
    \subfigure[]{\includegraphics[width=0.495\textwidth]{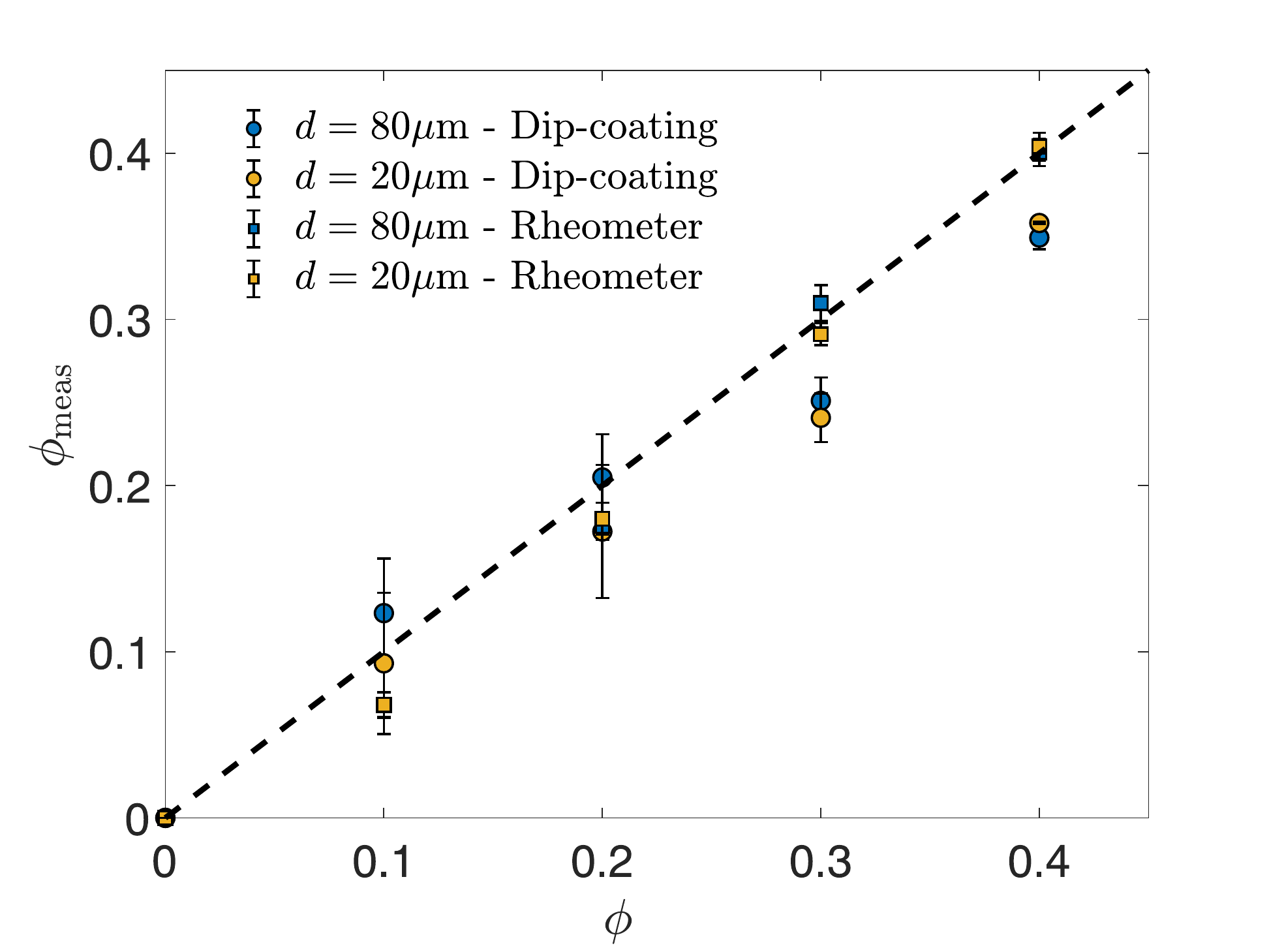}}
    \caption{(a) Relative shear viscosity $\eta_{\rm r}$ of suspensions of particle diameter
        $d = 20\,\mu{\rm m}$ and $d = 80\,\mu{\rm m}$ as a function of the particle volume fraction in the bath $\phi$ (circles). $\eta_{\rm r}$ is estimated from the thickness of the coating film using equation~(\ref{eq:LLD_effVisc}). The shear viscosity measured with the rheometer is also reported (squares). The dashed line is the Maron-Pierce correlation [equation (\ref{eq:MP})]. (b) Comparison between the particle volume fraction of the suspensions in the bath, $\phi$, and the value obtained from the viscosity $\phi_{\rm meas}$, measured either by dip-coating or rheometer. The dashed line has a slope $1$.}
    \label{fig:Figure_4}
\end{figure}

We observe that for both particle diameters ($20\,\mu{\rm m}$ and $80\,\mu{\rm m}$), the experimental data collapse onto the LLD law when $h\gtrsim d$ (figures~\ref{fgr:Figure_3}(c)-(d)). A similar observation can also be done for other sizes of particles used in this study. Figure \ref{fig:Figure_4}(a) reports the relative effective viscosity of the suspension, $\eta_{\rm r}=\eta(\phi)/\eta_0$ obtained through this approach. The evolution is similar for both particle sizes: at small enough volume fractions  ($\phi \lesssim 0.2$), the viscosity follows the Mason-Pierce correlation [equation (\ref{eq:MP})] although it is slightly smaller than the viscosity of the suspension in the bath. 

At larger volume fractions, the viscosity obtained by fitting the experimental data with the LLD law is systematically lower than the viscosity of the suspension in the bath. The larger difference in viscosity observed for larger volume fractions is due to the nonlinearity of the evolution of $\eta_{\rm r}$ with $\phi$. The difference between the actual volume fraction of the suspension in the bath and the estimated volume fraction of the coating film is reported in figure \ref{fig:Figure_4}(b). The difference is approximatively equal to $\Delta\phi = 0.008, 0.012, 0.056, 0.049$ for  $\phi=0.1, 0.2, 0.3, 0.4$, and thus shows a relative variation of  $\Delta\phi/\phi=8\%,\, 6\%, 18\%$ and $12\%$. The decrease in particle volume fraction in the coating film has been previously reported, yet it was of smaller magnitude \cite[][]{palma2019dip}. This small variation in volume fraction could be an effect of "self-filtration" due to the abrupt change in the flow at the stagnation point. This effect has been investigated by \cite{kulkarni2010particle} for a gravity-driven flow of dense suspensions  ($\phi>0.5$) through a wide aperture. Here, the stagnation point and the dynamic meniscus also play the role of an aperture, with one solid boundary and one deformable boundary imposed by the air-liquid interface, so that a similar "self-filtration" effect can be expected. Note that the resulting difference between the coating thickness and the predicted value by the LLD law and the Maron-Pierce correlation is small (about $10\%$) and was already visible in previous measurements \cite[][]{gans2019dip}. We insist on this point to stress the difference between the viscosity decrease due to the self-filtration and the viscosity decrease to polydispersity, which can be of similar magnitude.
In the following section, we consider the role of a bimodal distribution of particle size on the coating film.


\section{Coating of bidisperse suspensions} \label{sec:bi}

\subsection{General observations}

\begin{figure}
\centering
\includegraphics[width=1\textwidth]{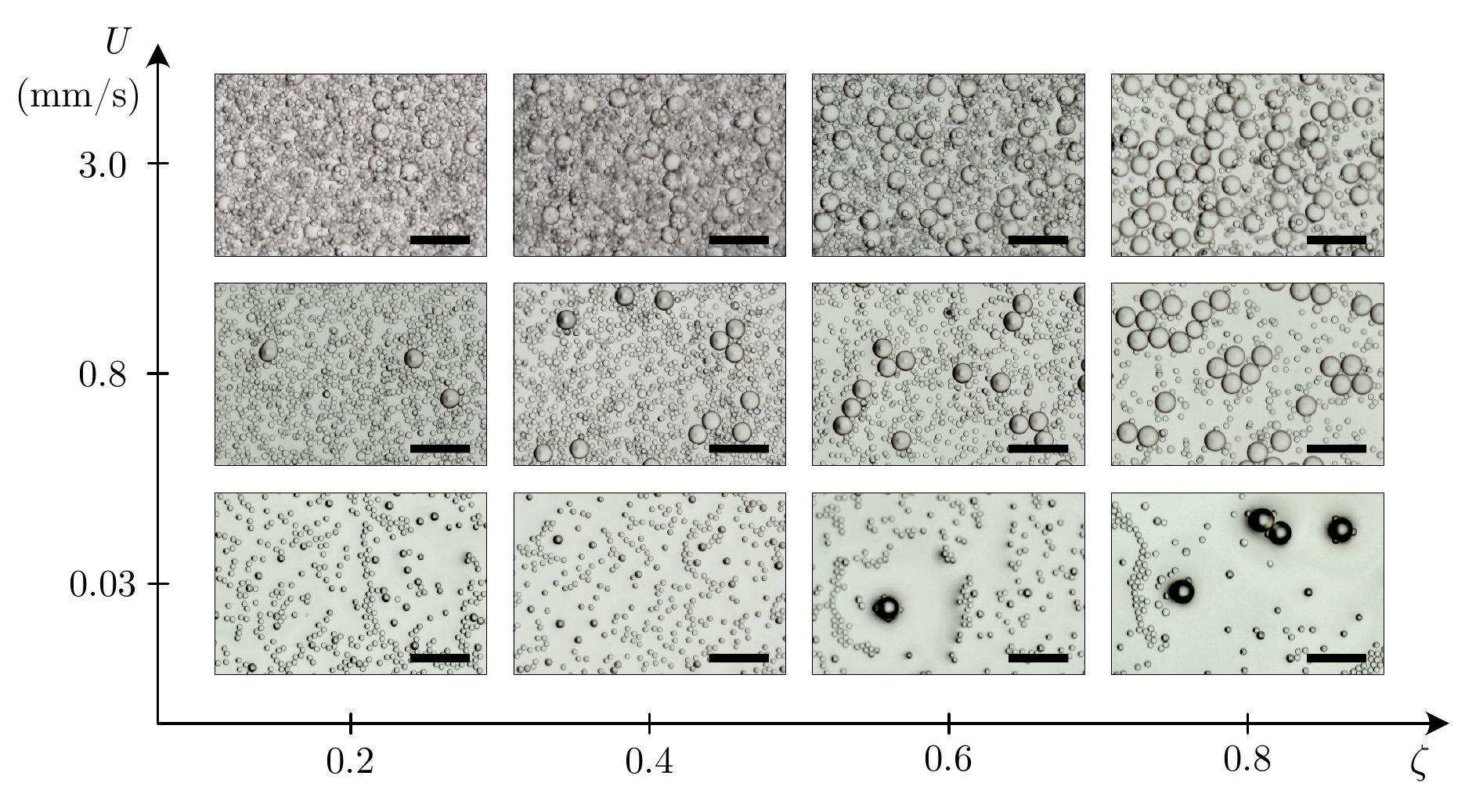}
  \caption{Examples of coating films observed for bidisperse suspensions 
      of particles diameters $d_{\rm S}=20\,\mu{\rm m}$ and $d_{\rm L}=80\,\mu{\rm m}$ 
      (size ratio $\delta=4$), at a volume fraction $\phi=0.2$ 
      and different volume ratios of large particles ($\zeta=0.2,\,0.4,\,0.6,$ and $0.8$) 
      for increasing withdrawal velocities $U$.
      The size of the scale bars is 250 $\mu m$.
  }
  \label{fgr:Figure_5}
\end{figure}

In this section, we consider suspensions of particles having a bimodal size distribution: small particles of diameter $d_{\rm S}$ and large particles of diameter $d_{\rm L}$. The composition of the solid phase is defined by the volume ratio of large particles: $\zeta=V_{\rm L}/(V_{\rm L}+V_{\rm S})$, where $V_{\rm L}$ and $V_{\rm S}$ are the volume of large and small particles in the suspension, respectively. Figure \ref{fgr:Figure_5} shows examples of typical coating patterns observed for different withdrawal velocity $U$ and for different compositions of the solid phase, with $d_{\rm S}=20\,\mu{\rm m}$ and $d_{\rm L}=80\,\mu{\rm m}$. The solid volume fraction is kept constant and equal to $\phi=0.2$. We observe that the composition of the coating film changes drastically in terms of particle size distribution and depends both on $U$ and $\zeta$. For instance, for a balanced composition ($\zeta=0.4$) and a low withdrawal velocity, only the small particles are entrained in the coating film. For a given value of $\zeta$, the number of large particles increases with the withdrawal velocity, and thus the thickness $h$ of the film. We observe the same behavior for all compositions of $\zeta$ considered here.

A deficit of large particles in the coating film is observed for low or moderate withdrawal velocity. Indeed, when the thickness at the stagnation point $h^*$ is smaller than the particle radius, the particles are filtered out of the film \cite[][]{sauret2019capillary}. It remains unclear if the three regimes, which reported for monodisperse suspensions ("liquid only", "heterogeneous films", and "effective viscosity"), still hold for polydisperse suspensions.

The experiments reported in Figure \ref{fgr:Figure_1} and Figure \ref{fgr:Figure_5} suggest that at low velocity and small enough volume fraction, a first coating regime is observed. Within this regime, the coating film does not include any particles and corresponds to the "liquid-only" regime observed for monodisperse suspensions \cite[][]{gans2019dip}. At large withdrawal velocities, multiple layers of particles are visible, and the composition of the coating film is comparable to the composition of the suspension bath. This regime, which corresponds to an effective viscosity regime, is studied in detail in section \ref{sec:thick}. The regime in-between those two regimes, the heterogeneous regime, is more complex for bidisperse suspensions. At low withdrawal velocity, the volume ratio of large particles $\zeta$ in the coating film is smaller than that in the bath, \textit{i.e.} the film mostly contains small particles, and only a few large particles can be seen. Here, the small particles reach their effective viscosity regime while the large particles only start to be entrained in the film. Increasing the withdrawal velocity leads to an increase in the number of large particles in the coating film. We discuss this heterogeneous regime in Section \ref{sec:heterogeneous}.

\subsection{Effective viscosity regime} \label{sec:thick}

\subsubsection{Experimental results}

\begin{figure}
\centering
\subfigure[]{\includegraphics[width=0.495\textwidth]{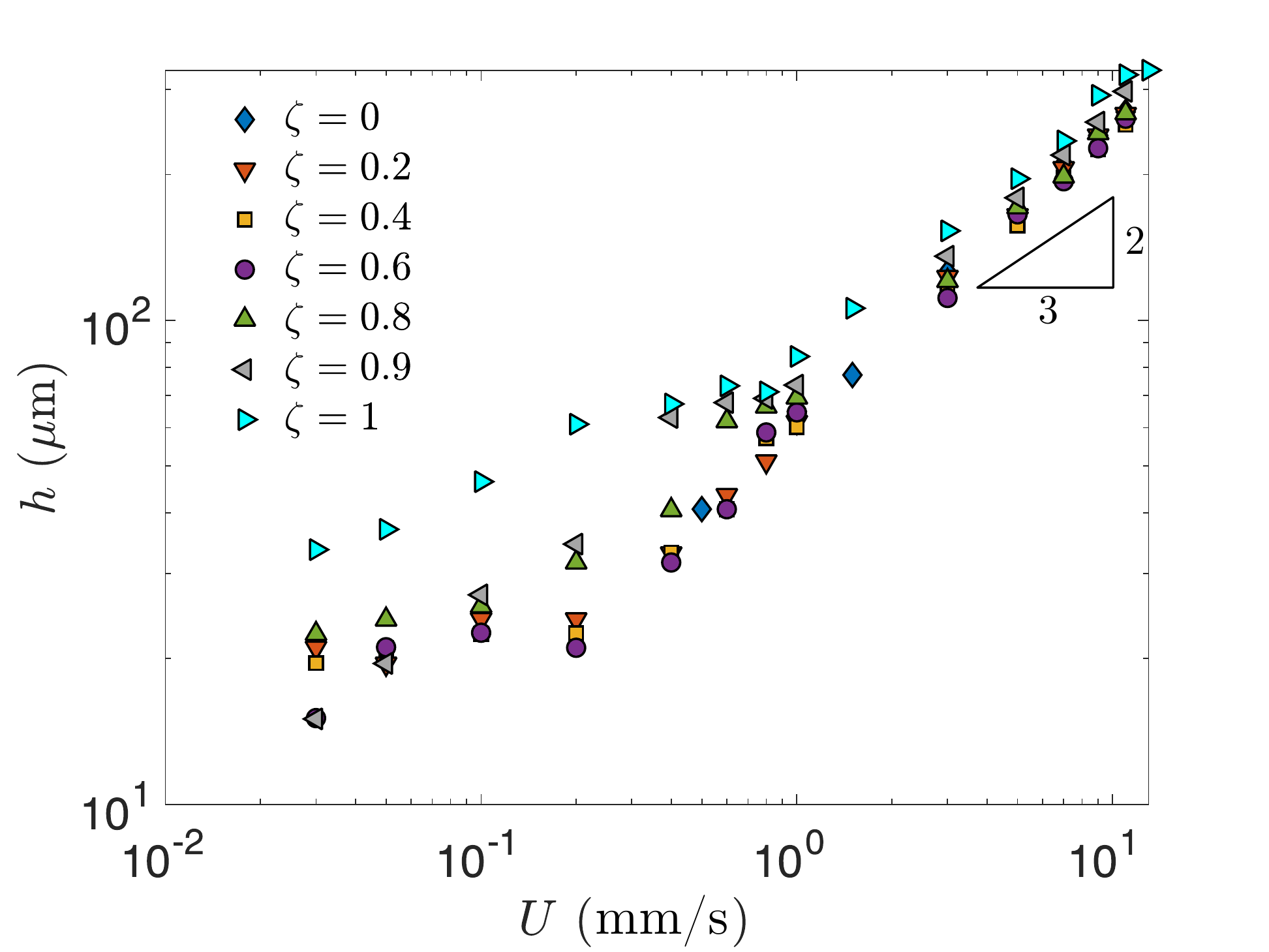}}
\subfigure[]{\includegraphics[width=0.495\textwidth]{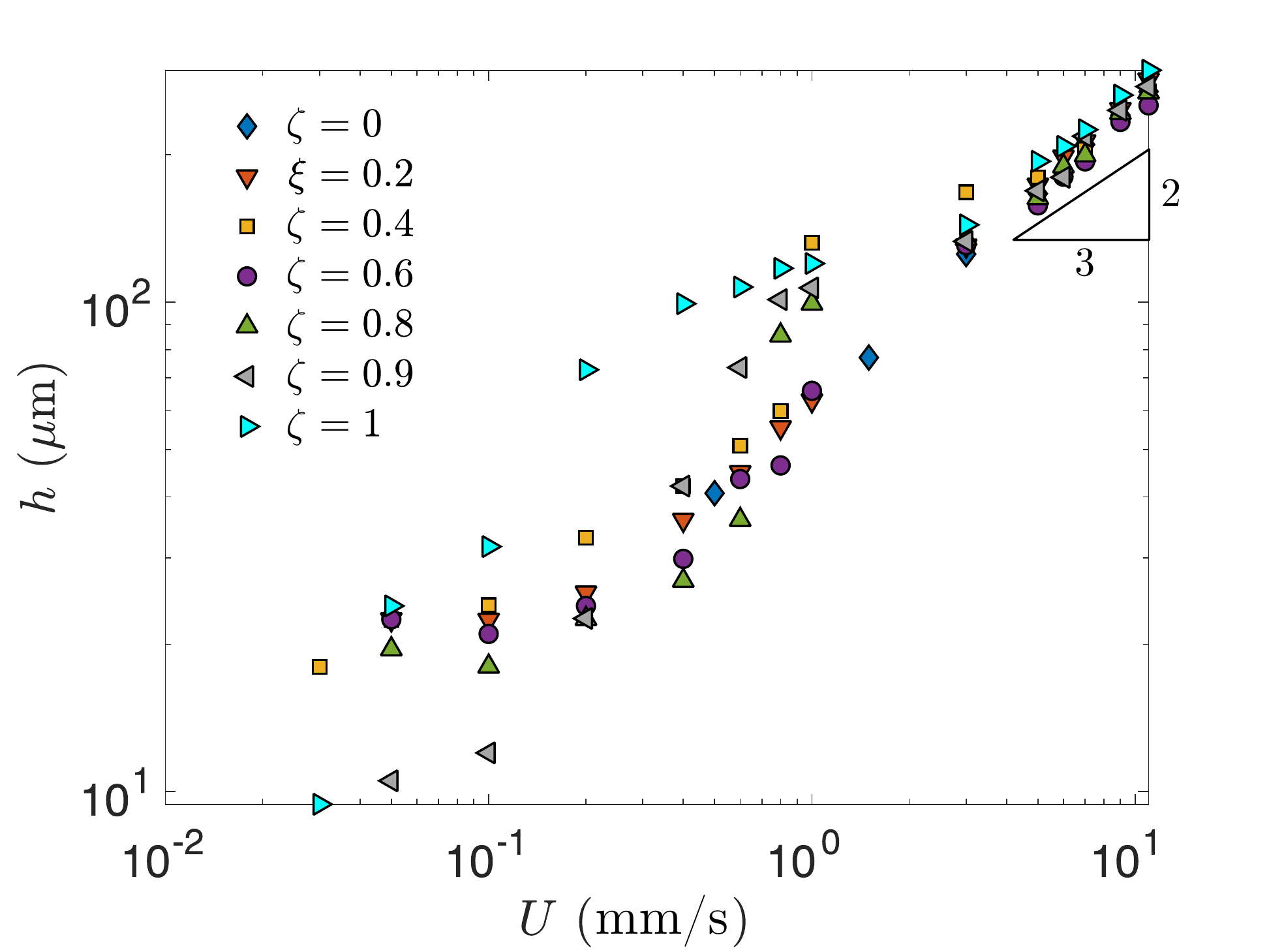}}
\subfigure[]{\includegraphics[width=0.495\textwidth]{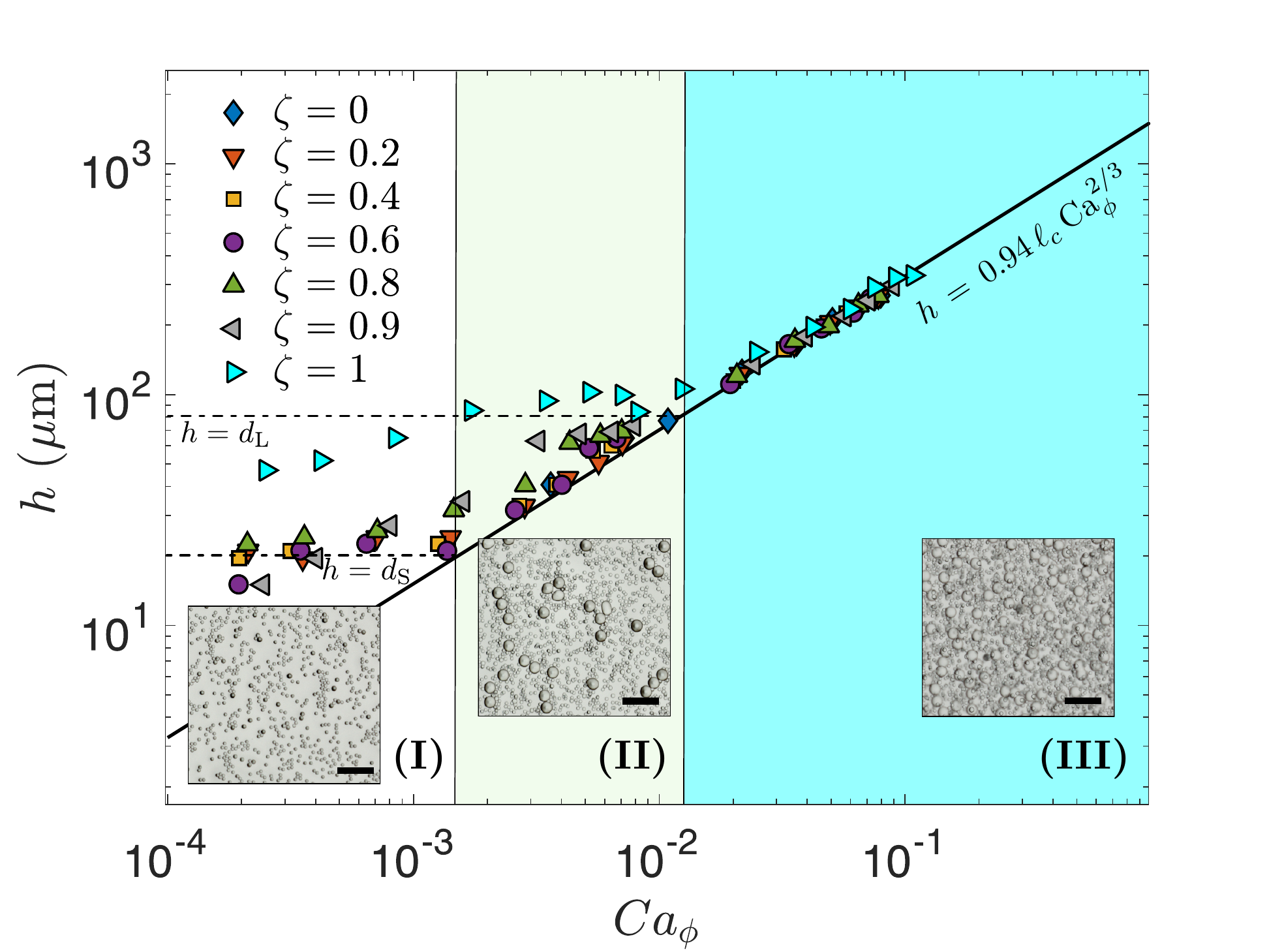}}
\subfigure[]{\includegraphics[width=0.495\textwidth]{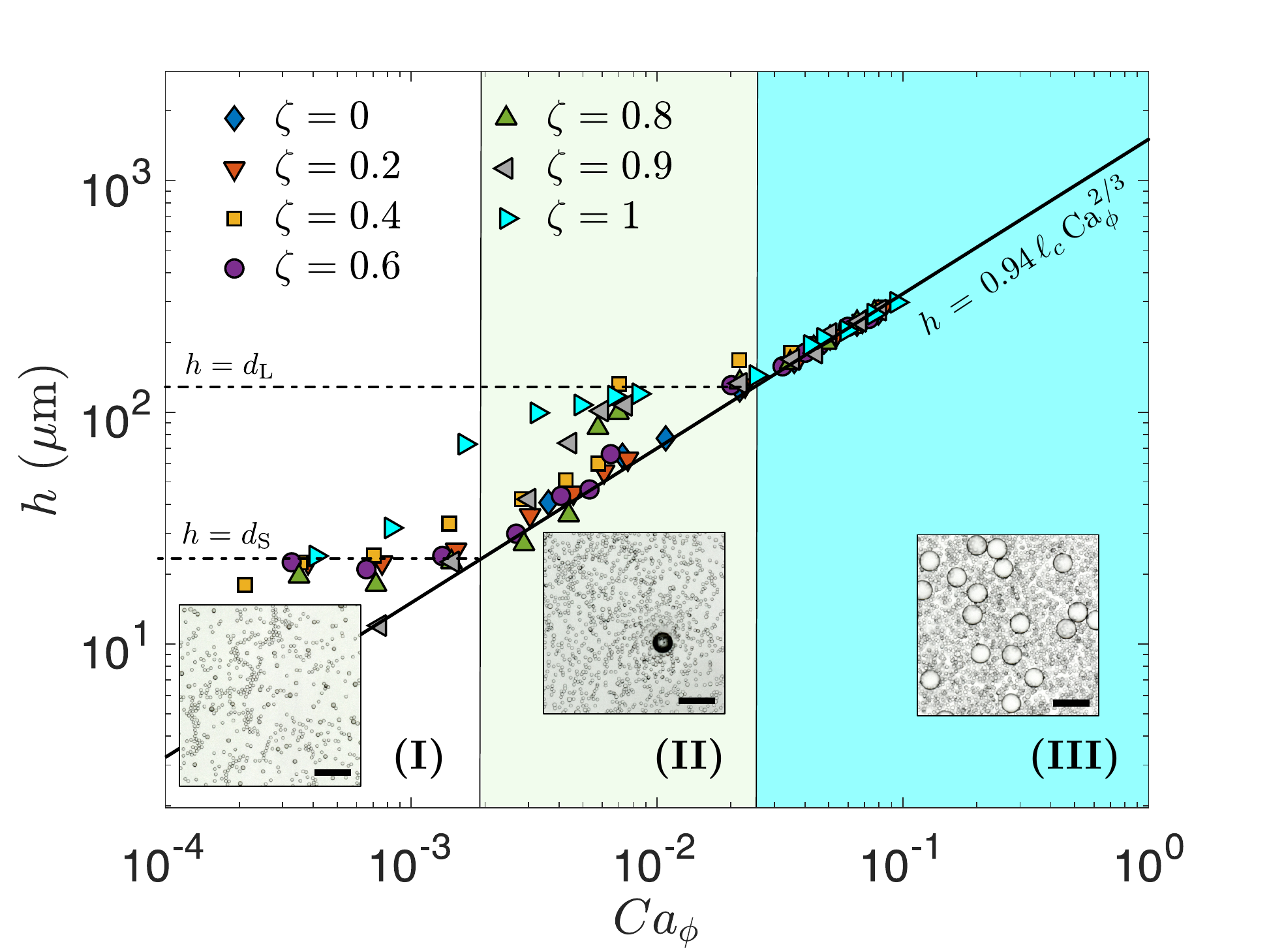}}
  \caption{Thickness of the coating film versus (a)-(b) the withdrawal velocity $U$ and 
      (c)-(d) the effective capillary number ${\rm Ca}_\phi$ for bidisperse suspensions with
      particles of diameter
      (a)-(c) $d_{\rm S} = 20\,\mu{\rm m}$ and $d_{\rm L} = 80\,\mu{\rm m}$ ($\delta=4$); 
      (b)-(d) $d_{\rm S} = 20\,\mu{\rm m}$ and $d_{\rm L} = 140\,\mu{\rm m}$ ($\delta=7$).
      The volume fraction is $\phi = 20\%$.
      On each figure, we vary the volume ratio of large particles $\zeta$ from $0$ to $1$.
      In (c) and (d) the two horizontal dashed line respectively corresponds 
      to a coating film thickness equals to the particle diameter, $h=d_{\rm S}$ and $h=d_{\rm L}$.
      The continuous thick line is the LLD law, where the viscosity is considered as a fitting parameter.
      We observe three regimes:
      (I) is the regime where the film thickness is more or less constant and similar to $d_{\rm S}$.
      (II) is the regime where the film is primarily composed of small particles, 
      and the number of large particles depends on the withdrawal velocity.
      (III) is the thick-film regime where the coating film is thicker than the diameter of the large particles, and the composition of the coating film is similar to that of the bath. Inset pictures show the coating film in these three regimes. The size of the scale bars is $250 \mu\rm{m}$.
  }
  \label{fgr:Figure_6}
\end{figure}

\begin{figure}
\centering
\subfigure[]{\includegraphics[width=0.495\textwidth]{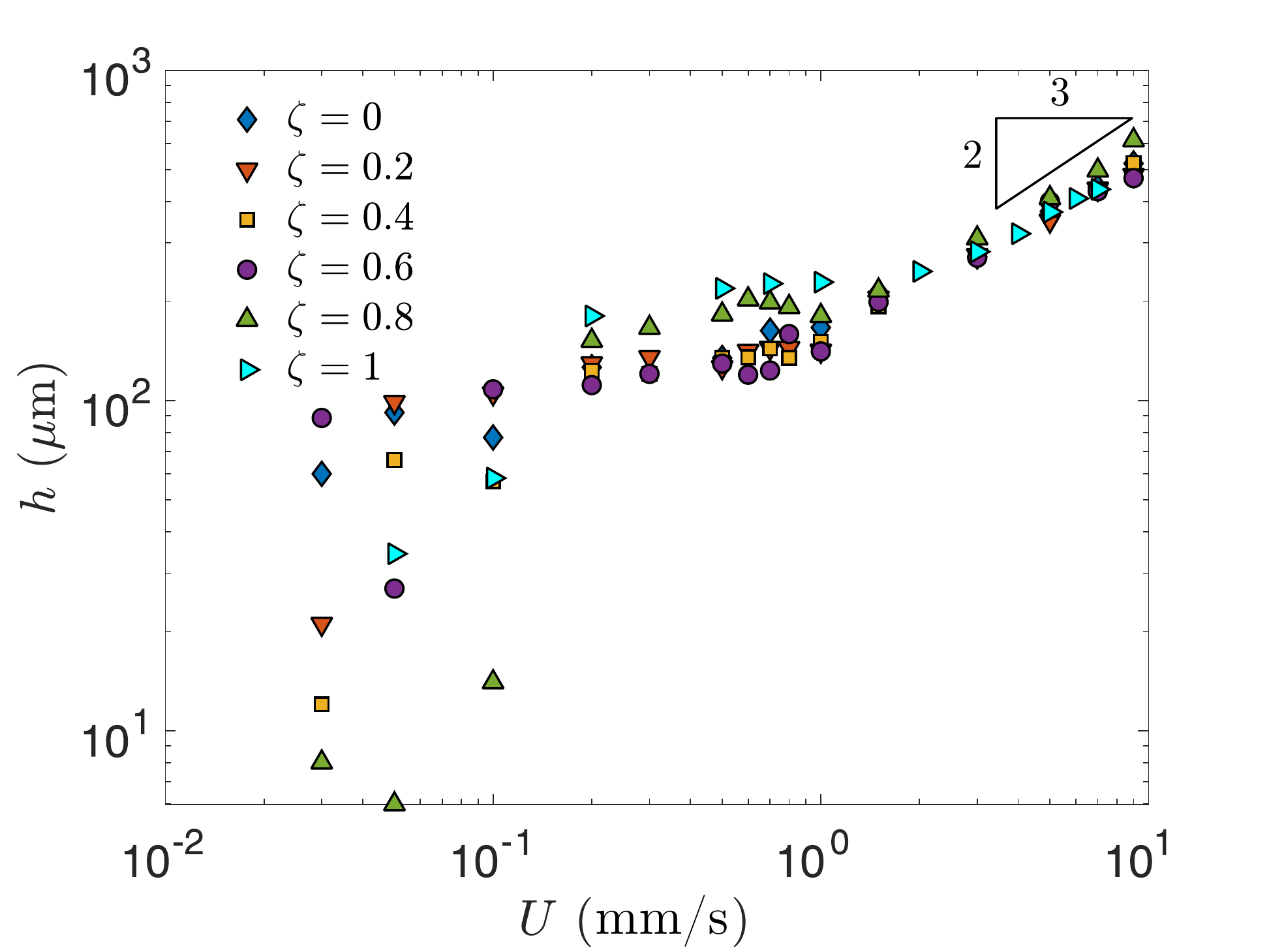}}
\subfigure[]{\includegraphics[width=0.495\textwidth]{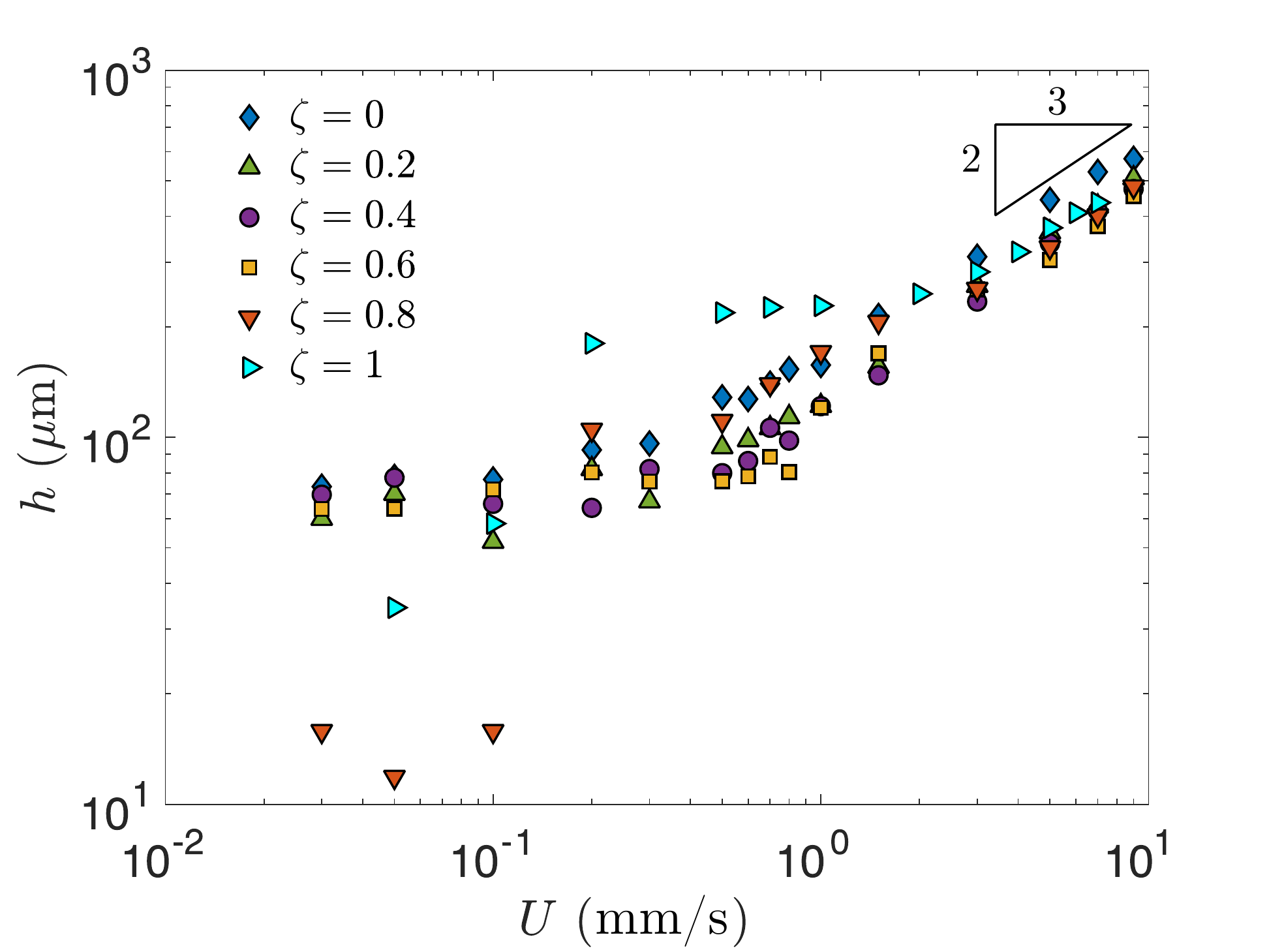}}
\subfigure[]{\includegraphics[width=0.495\textwidth]{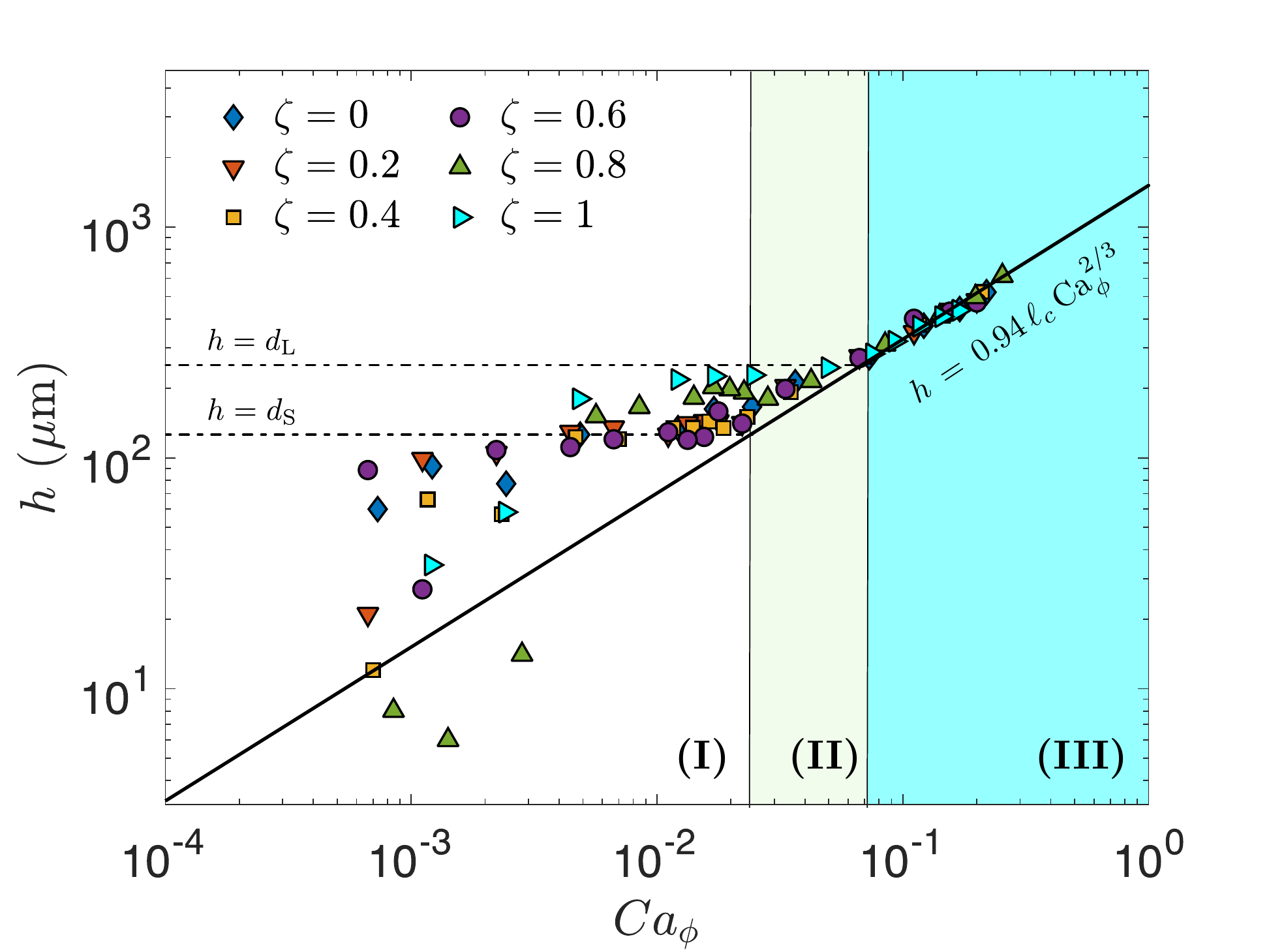}}
\subfigure[]{\includegraphics[width=0.495\textwidth]{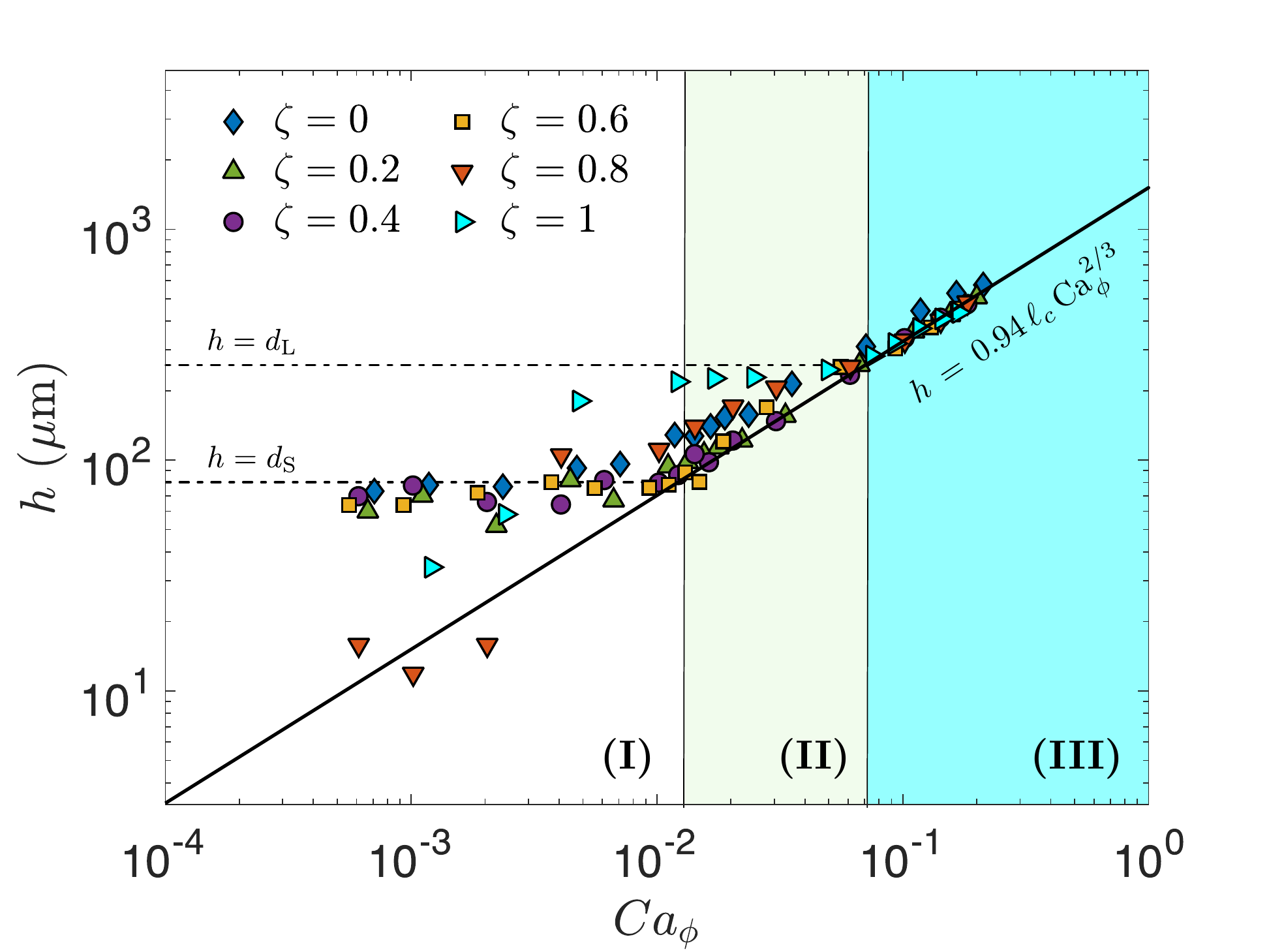}}
\caption{Thickness of the coating film versus (a)-(b) the withdrawal velocity $U$,
    and (c)-(d) the effective capillary number ${\rm Ca}_\phi$ for bidisperse suspensions with
    particles of diameter 
    (a)-(c) $d_{\rm S} = 140\,\mu{\rm m}$ and $d_{\rm L} = 250\,\mu{\rm m}$ ($\delta=1.786$);
    (b)-(d) $d_{\rm S} =  80\,\mu{\rm m}$ and $d_{\rm L} = 250\,\mu{\rm m}$ ($\delta=3.125$).
    The volume fraction is $\phi = 40\%$.
    In figures (c) and (d) the two horizontal dashed line corresponds to a coating film thickness
    equals to the particle diameters, respectively $h=d_{\rm S}$ and $h=d_{\rm L}$.
    The continuous thick line is the LLD law, where the viscosity is considered as a fitting parameter.
}
  \label{fgr:Figure_7}
\end{figure}

We measure the thickness of the coating film for the bidisperse suspension shown in figure \ref{fgr:Figure_5} varying the withdrawal velocity and the volume ratio of large particles $\zeta$. Figure \ref{fgr:Figure_6}(a) shows the thickness of the coating film as a function of the withdrawal velocity. In the regime of fast withdrawal ($U \gtrsim 2\,{\rm mm/s}$), monodisperse and bidisperse suspensions follows a common power-law $h \propto U^{2/3}$. This observation suggests that an effective viscosity can also be extracted for a thick enough film of bidisperse suspension. We perform an analysis similar to the one used for the monodisperse suspensions in section \ref{sec:mono} and fit the thickness $h$ to the LLD law [equation (\ref{eq:LLD_effVisc})], where the effective viscosity $\eta(\phi)$ is the fitting parameter. The rescaling is shown in figure \ref{fgr:Figure_6}(c). It demonstrates that the coating film is in the effective viscosity regime, provided that the film is thicker than the diameter of the large particles, $h \geq d_{\rm L}$.

The situation is nevertheless more complex than for monodisperse suspensions. Indeed, the threshold to the effective viscosity regime seems to depend on the volume ratio of large particles $\zeta$. For small $\zeta$ (for instance $\zeta=0.2$ in figure \ref{fgr:Figure_6}(c)) the thickness of the film follows fairly well the LLD law as soon as $h \geq d_{\rm S}$. Indeed, small values of $\zeta$ mean that the volume of large particles is small compared to the volume of small particles. Therefore, the large particles do not contribute significantly to the viscosity of the suspension. Note that although the prediction of the LLD law is reasonably good when $d_{\rm S} < h < d_{\rm L}$, the composition of the coating film is different from the composition of the suspension bath with a deficit in large particles as we shall see in section \ref{sec:heterogeneous}. For large values of $\zeta$ (for instance, $\zeta=0.8$ or $0.9$ in figure \ref{fgr:Figure_6}(c)), the LLD law is recovered only for $h \geq d_{\rm L}$. In this case, the large particles are the main contributor to the viscosity of the suspension. Thus, recovering the LLD law requires the coating film to be thick enough ($h \geq d_{\rm L}$) so that it can allow most of the particles to be entrained in the film.

The same observation can be made with another combination of particle sizes: Figure \ref{fgr:Figure_6}(b) shows the case of a suspension with $d_{\rm L}=140\,\mu{\rm m}$ and $d_{\rm S}=20\,\mu{\rm m}$ particles. In this case, the effective viscosity regime following the LLD law is also recovered for
$h\geq d_{\rm L} = 140\,\mu{\rm m}$ [figure \ref{fgr:Figure_6}(d)]. Here, a similar evolution than the one reported in figures \ref{fgr:Figure_6}(a) and \ref{fgr:Figure_6}(c) is observed.

We also considered a larger volume fraction: $\phi=40\%$ of $d_{\rm L}=250\,\mu{\rm m}$/$d_{\rm S}=140\,\mu{\rm m}$ [figure \ref{fgr:Figure_7}(a)] and $d_{\rm L}=250\,\mu{\rm m}$/$d_{\rm S}=80\,\mu{\rm m}$ [figure \ref{fgr:Figure_7}(b)]. Again, the coating thickness $h$ follows the LLD law with an effective capillary number ${\rm Ca}_\phi$, where the viscosity of the bidisperse suspension is still considered as a fitting parameter. A similar behavior than the one reported for suspensions at $\phi=20\%$ is observed:  the effective viscosity regime starts at $h \geq d_{\rm L}$, and the transition from the heterogeneous film to the effective viscosity regime is smoother  for small fraction of large particles $\zeta$ [figures \ref{fgr:Figure_7}(c)-(d)]. Besides, since the particles used here are larger than the ones used in the $\phi=20\%$ case, we are also able to see the liquid-only regime, where barely any particles are entrained. This regime is observed at small values of ${\rm Ca}_\phi$, and thus small $h$ (data on the bottom left corner indicated as (I) in figures \ref{fgr:Figure_7}(c)-(d)).

 \subsubsection{Effective viscosity of bidisperse suspensions}
 
Although the LLD law is recovered for bidisperse suspensions when $h\geq d_{\rm L}$, figures \ref{fgr:Figure_6}(a)-(b) and \ref{fgr:Figure_7}(a)-(b) show that for a given value of $U$ and $\phi$, a change in the volume ratio of large particles $\zeta$ leads to a change in the film thickness. This observation is consistent with the influence of the composition of the solid phase on the viscosity: a change in $\delta$ or $\zeta$ causes a change in viscosity, hence a change in film thickness~\cite[][]{shapiro1992random,gamonpilas2016shear,thievenaz2021droplet}.

\begin{figure}
\centering
\subfigure[]{\includegraphics[width=0.495\textwidth]{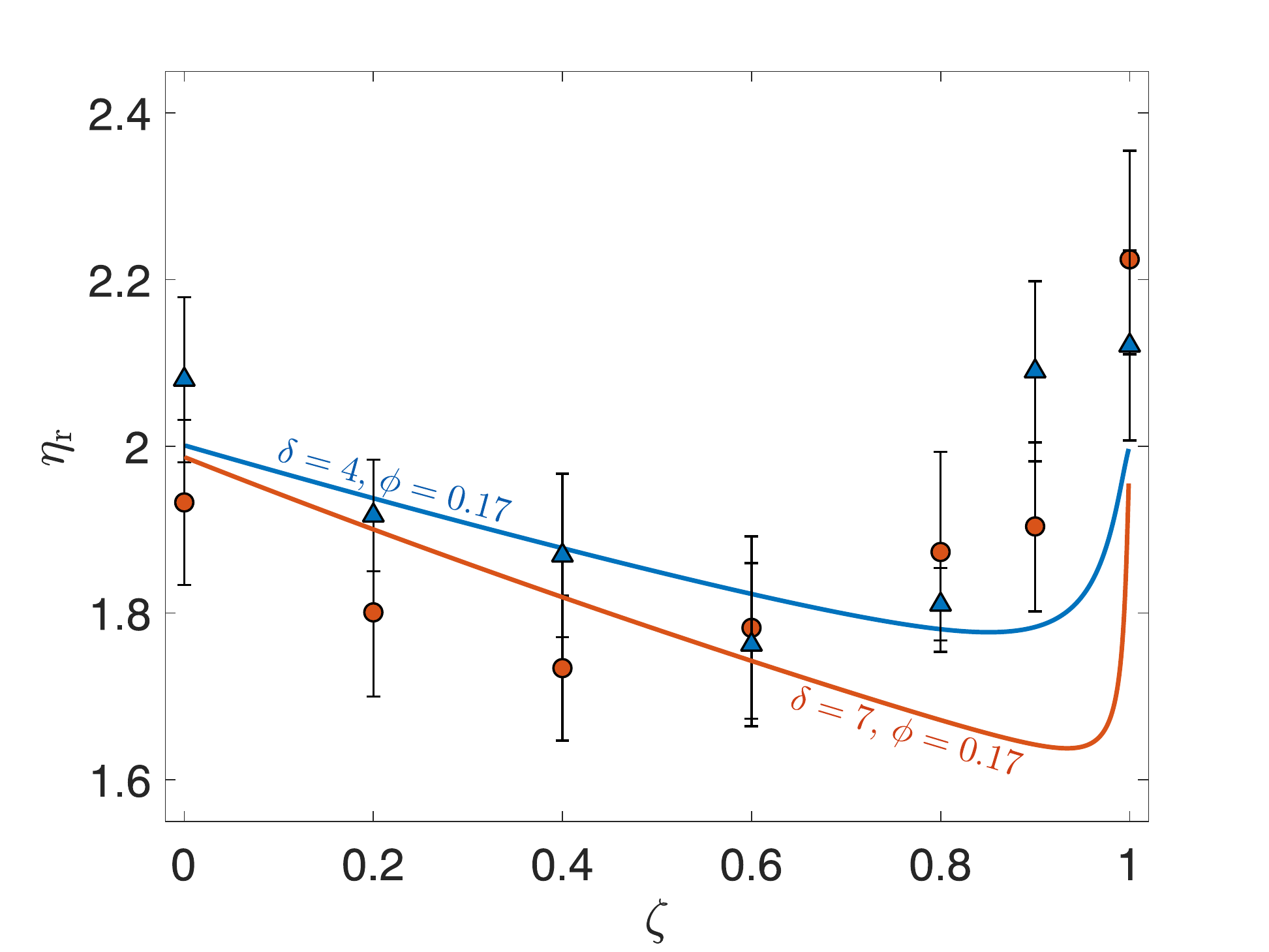}}
\subfigure[]{\includegraphics[width=0.495\textwidth]{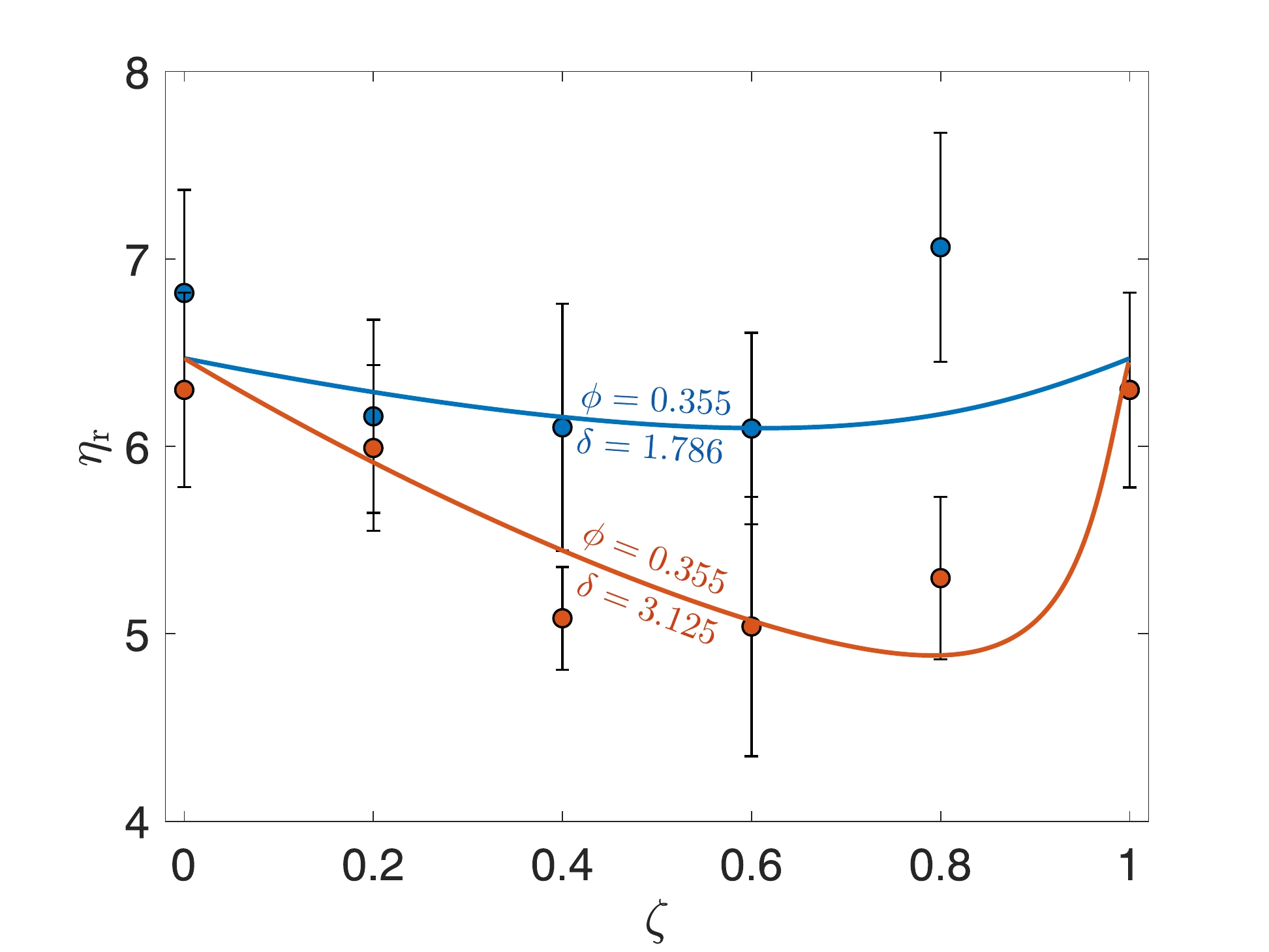}}
\caption{Relative shear viscosity $\eta_{\rm r}$ versus the volume fraction of large particles
    $\zeta$ for a suspension containing 
    (a) $\phi=20\%$ for particle size ratio $\delta=4$ (blue) and $7$ (red) 
    and using $\phi_{\rm film}=17\%$ in the coating film; 
    (b) $\phi=40\%$ for particle size ratio $\delta=1.786$ (blue) $3.125$ (red)
    and using $\phi_{\rm film}=35.5\%$ in the coating film,.
    The symbols show data obtained by the best fit to the LLD law.
    The lines show to the viscosity predicted by the Maron-Pierce correlation [equation (\ref{eq:MP})]
    using the maximum packing fraction given by equation(\ref{eq:model_phic}).}
  \label{fgr:Figure_8}
\end{figure}

The effective viscosity $\eta(\phi,\,\delta,\,\zeta)$ is derived by fitting the experimental data to the LLD law in the effective viscosity regime. Figure \ref{fgr:Figure_8}(a) (resp. \ref{fgr:Figure_8}(b)) reports the relative viscosity $\eta_{\rm r}=\eta/\eta_0$ as a function of the volume ratio of large particles $\zeta$  for the experiments presented in figures \ref{fgr:Figure_6}(c)-(d) (resp. figures \ref{fgr:Figure_7}(c)-(d)). In figure \ref{fgr:Figure_8}(a), the volume fraction in the bath is $\phi=20\%$ and the size of the particles are $20\,\mu{\rm m}/80\,\mu{\rm m}$ and $20\,\mu{\rm m}/ 140\,\mu{\rm m}$. In figure \ref{fgr:Figure_8}(b), the volume fraction in the bath is $\phi=40\%$ and the size of the particles are $140\,\mu{\rm m}/250\,\mu{\rm m}$ and $80\,\mu{\rm m}/250\,\mu{\rm m}$. Between the two monodisperse cases, the relative viscosity $\eta_{\rm r}$ as a function of $\zeta$ shows a parabolic curve, reaching its minimum around $\zeta \simeq 0.4-0.6$. This canonical behavior of bidisperse suspensions is due to the higher compacity of bidisperse packings~\cite[see \textit{e.g.}][]{pednekar2018bidisperse}. The difference in viscosity observed between the two monodisperse cases, at $\zeta=0$ and $\zeta=1$, arises from the size variance of the particles. Although the suspensions used are monodisperse down to a certain level  (see the size measurement in Appendix), a small amount of polydispersity is unavoidable and therefore, the maximum packing fraction for these two distributions of particles is slightly different, about $5-10\%$ here.

For a bidisperse suspension in the effective viscosity regime ($h > d_{\rm L}$), the proportions of small and large particles are expected to be similar in the film and in the bath. The evolution of the viscosity can be modeled by calculating the maximal packing fraction $\phi_{\rm m}(\delta,\,\zeta)$ of a bidisperse sphere packing of the same composition and then substituting it in the Maron-Pierce correlation [equation (\ref{eq:MP})]. To compute the maximal packing fraction, we adapt the model of \cite{ouchiyama1984porosity}. It consists in computing the local compacity around each size of particles and averaging it over the size distribution. The model is simplified to consider here a bimodal size distribution. The number fractions of small $N_{\rm S}$ and large $N_{\rm L}$ particles are defined as
\begin{equation}
    N_{\rm S} = \frac{(1-\zeta)\, \delta^3}{(1-\zeta)\, \delta^3 + \zeta}
    \quad {\rm and } \quad
    N_{\rm L} = \frac{\zeta}{(1-\zeta)\, \delta^3} \,N_{\rm S},
    \label{eq:model_N}
\end{equation}
respectively, and $\tilde{d_\mathrm{S}} = d_\mathrm{S}/\delta$ and $\tilde{d_\mathrm{L}} = d_\mathrm{L}/\delta$ are the reduced sizes given by
\begin{equation}
    \tilde{d_\mathrm{S}} = \frac{(1-\zeta)\,\delta^3 + \zeta }{(1-\zeta)\,\delta^3 + \zeta\delta}
    \quad {\rm and } \quad
    \tilde{d_\mathrm{L}} = \delta\,\tilde{d_\mathrm{S}}.
    \label{eq:model_d}
\end{equation}
The maximum packing fraction of the bidisperse packing is then given by
\begin{equation}
    \phi_{\rm m}(\delta,\zeta) = \frac{N_{\rm S}\, \tilde{d_\mathrm{S}}^3 + N_{\rm L} \tilde{d_\mathrm{L}}^3}
    {(N_{\rm S}/\Gamma)(\tilde{d_\mathrm{S}}+1)^3 + N_{\rm L} \left( (\tilde{d_\mathrm{L}}-1)^3 + \left[(\tilde{d_\mathrm{L}}+1)^3 - (\tilde{d_\mathrm{L}}-1)^3 \right]/\Gamma \right)},
    \label{eq:model_phic}
\end{equation}
where $\Gamma$ denotes the average number of particles in the vicinity of a given particle and is equal to
\begin{equation}
    \Gamma = 1 + \frac{4}{13} (8\phi_{m,0}-1)\,
    \frac{N_{\rm S}\,(\tilde{d_\mathrm{S}} +1)^2 \left( 1 - \frac{3}{8} \frac{1}{\tilde{d_\mathrm{S}}+1} \right)
        + N_{\rm L} (\tilde{d_\mathrm{L}}+1)^2 \left( 1 - \frac{3}{8} \frac{1}{\tilde{d_\mathrm{L}}+1} \right)}
        {N_{\rm S}\,\tilde{d_\mathrm{S}}^3 + N_{\rm L} \left[ \tilde{d_\mathrm{L}}^3 - (\tilde{d_\mathrm{L}}-1)^3\right]}
    \label{eq:model_gamma}
\end{equation}
Here, $\phi_{m,0}$ is the maximum solid fraction in a monodisperse packing,  which we estimated through our rheometer measurements at $\phi_{m,0} \simeq 58\%$. We then compute $\phi_{\rm m}$ using equation~(\ref{eq:model_phic}) and obtain the viscosity through the Maron-Pierce correlation given by equation~(\ref{eq:MP}). This approach has been previously used to describe the viscosity of a bidisperse suspension in an oscillating plane Couette flow \cite[][]{gondret1997dynamic},  or the detachment of drops of bidisperse suspensions \cite[][]{thievenaz2021droplet}.

Figures \ref{fgr:Figure_8}(a)-(b) report the viscosity measured by fitting the dip-coating results to the LLD law, and compare it to the predictions of equations (\ref{eq:MP}) and (\ref{eq:model_phic}). These predictions match our experiments well, proving that the bidisperse suspensions behave like an effective viscous fluid. Achieving this good match requires that we use a volume fraction $\phi_{\rm film}$ slightly smaller than the volume fraction in the suspension bath $\phi$ as $\phi_{\rm film} \simeq 17\%$ for $\phi=20\%$ and $\phi_{\rm film} \simeq 35.5\%$ for $\phi=40\%$. This discrepancy is consistent with the self-filtration effect \cite[][]{kulkarni2010particle} and the observations made in the previous section for monodisperse suspensions as there are fewer particles in the film than in the bath, regardless of their sizes.

The comparisons between the viscosity obtained from the LLD law and the model show that at small size ratios ($\delta = d_{\rm L}/d_{\rm S}$), the viscosity is well predicted over the whole range of $\zeta$. However, at larger values of $\delta$, the model usually fails for $\zeta>60\%$, \textit{i.e.}, when large particles dominates. This is explicit in figure \ref{fgr:Figure_8}(a). The same failure of the model has been observed in other configurations \cite[][]{gondret1997dynamic,thievenaz2021pinch}. Therefore, the mismatch between experimental results and the prediction roots from the limitation of the model for the viscosity (equation~\ref{eq:MP} and \ref{eq:model_phic}) and not a problem specific to dip-coating.

In summary, for a bidisperse suspension, the effective viscosity regime is observed for a coating thickness larger than the diameter of the largest particles $h > d_{\rm L}$. This condition can be expressed in terms of capillary number associated with the interstitial fluid:

\begin{equation}
    {{\rm Ca}_0}^* \geq 1.09\,\frac{\eta_0}{\eta(\phi)}\,\left(\frac{d_{\rm L}}{\ell_c}\right)^{3/2},
    \quad {\rm with} \quad {\rm Ca}_0=\eta_0\,U/\gamma
\end{equation}
or associated to the capillary number based on the effective viscosity of the bidisperse suspension
\begin{equation}
    {{\rm Ca}_\phi}^* \geq 1.09\,\left(\frac{d_{\rm L}}{\ell_c}\right)^{3/2}, 
    \quad {\rm with} \quad {\rm Ca}_\phi=\frac{\eta(\phi,\delta,\zeta)\,U}{\gamma}
\end{equation}
This threshold is similar to the case of monodisperse suspension \cite[][]{gans2019dip,palma2019dip} but only depends on the diameter of the large particle. In the effective viscosity regime, the thickness of the coating film can be estimated using the LLD law with a capillary number based on the effective viscosity of the bidisperse suspension:

\begin{equation}
    h=0.94\,\ell_c\,{{\rm Ca}_\phi}^{2/3}
\end{equation}

For a given solid volume fraction, bidisperse suspensions are less viscous than monodisperse suspensions. This decrease in viscosity is more pronounced for a large difference in the particle size ratio, that is, when $\delta$ is high (see figure \ref{fgr:Figure_8}(a)-(b)). Therefore, in the effective film regime, bidisperse suspensions yield thinner films than monodisperse suspensions for a given volume fraction $\phi$.

\subsection{Heterogeneous regime} \label{sec:heterogeneous}

 \subsubsection{Observations}
 
For withdrawal velocities $U$ leading to coating films thinner than the diameter of the large particles  ($h \lesssim d_{\rm L}$), the coating thickness does not follow the Landau-Levich-Derjaguin law anymore. This situation is observed in regions (I) and (II) in figures \ref{fgr:Figure_6}(c)-(d) and figures \ref{fgr:Figure_7}(c)-(d). This heterogeneous regime, which was already observed for monodisperse suspensions, is different depending on the range of sizes of the particles. The heterogeneous regime can be split into two regimes for bidisperse suspensions.

The first regime corresponds to the lowest capillary numbers, where the film is then mainly composed of small particles. Its thickness remains more or less constant and equal to $h \sim d_{\rm S}$ over a range of capillary number  (region (I) in figures \ref{fgr:Figure_6}(c)-(d) and \ref{fgr:Figure_7}(c)-(d)). A similar regime is observed for monodisperse suspensions and corresponds to a monolayer of small particles \cite[][]{gans2019dip,palma2019dip}. If the suspension is dilute enough and the particles are large enough, we can observe an extreme case where only the liquid is present in the coating film, without any particles entrained. This situation occurs at very small withdrawal velocities [left panel in figure \ref{fgr:Figure_1}]. It can also be seen when the suspension is primarily composed of large particles ($\zeta=0.8$ in figures \ref{fgr:Figure_7}(c)-(d)).
 
The second heterogeneous regime occurs at moderate capillary numbers, between the first heterogeneous regime and the effective viscosity regime. In this regime, corresponding to the region (II) in figures \ref{fgr:Figure_6}(c)-(d) and \ref{fgr:Figure_7}(c)-(d), the coating film is primarily composed of small particles but also contains some large particles. The number of entrained large particles increases continuously with the capillary number up to the effective viscosity regime when $h \geq d_{\rm L}$. For a small volume ratio of large particles, typically $\zeta=0.2$ or $0.4$, the LLD regime is reached earlier than for large $\zeta$. This observation can be rationalized by considering that the number of large particles remains small. Therefore the large particles do not contribute significantly to the effective viscosity of the suspension. The main challenge in predicting the threshold between the regimes lies in estimating the number of entrained particles. In the following subsection, we propose a filtration model that accounts, at first order, for the variation of composition in the coating film.
 

 \subsubsection{Discussion: Entrained Particle Distribution}

When $h<d_{\rm L}$, the interplay between the different length scales (different sizes of particles and film thickness) selects the particles entrained in the coating film. We present here a model that accounts for the variation in the composition of the coating film compared to the composition of the bath. We rely on the thickness of the coating film and the flow rate based on the Landau-Levich-Derjaguin theory of dip coating \cite[][]{levich1942dragging} at which we add the criterion given by equation (\ref{eq:condition_entrainment}), to set the minimum film thickness required for the entrainment of particles of diameter $d$ in the coating film. For a given particle size distribution in the bath, 
varying the withdrawal velocity and hence the thickness of the coating film will lead to a different particle size distribution in the coating film, as long as $h<d_{\rm L}$.

Consider the passage of particles from the bath to the film. We introduce $z$ as the coordinate parallel to the solid surface, $x$ as the coordinate perpendicular to the surface, and $s$ is the curvilinear coordinate that follows the meniscus (see figure \ref{fgr:Figure_1}(a)). The lubrication flow in the film reduces to the axial flow profile: $u_{z}^{*}(x, s)$. Then, the probability for a particle of diameter $d$ and position $(x, s)$ to be captured in the film is defined as $p_{c}^{*}(x, s, d)$. In addition, the particle-surface pair correlation function for a particle of diameter $d$ following a streamline passing through the point $(x, s)$ in the meniscus is defined as $g_{s}^{*}(x, s, d)$. This function describes the interaction between the surface and the particles. The equilibrium pair correlation may be needed to account for non-equilibrium effects for particles passing through the meniscus, for instance, the clustering of particles or interactions between them. From there, the flow rate of particles of diameter $d$ that enters the film, $Q_{\rm c}(d)$, is computed. The local flux is first integrated over a cross-section of the meniscus, perpendicular to the plate and passing through the stagnation curve defined as $x = h^*(s)$. $Q_{\rm c}(d)$ can be expressed as:
\begin{equation} \label{eq:model_1}
Q_{\rm c}(d)=\phi(d) \oint \int_{0}^{h^{*}(s)} u_{z}^{*}(x, s) p_{c}^{*}(x, s, d) g_{s}^{*}(x, s, d) {\rm d} x\, {\rm d} s,
\end{equation}
The general expression for the particle flow rate given by equation~(\ref{eq:model_1}) allows to account for complex solid surface geometries, such as fibers or textured plates \cite[][]{dincau2020entrainment,seiwert2011coating}, and general particle-surface correlations. Here, we can make some assumptions for the sake of simplification by considering the cross-section of a thin plate so that the $s$ dependence can be neglected during integration over the perimeter $P=2\,(W+e)$. A thin film assumption is also made in Cartesian coordinates ($x$, $z$):
\begin{equation} \label{eq:model_2}
    Q_{c}(d) = \phi(d) P 
    \int_{0}^{h^{*}} \,u_{z}^{*}(x)\, p_{c}^{*}(x, d)\, g_{s}^{*}(x, d)\, {\rm d} x.
\end{equation}
We further assume that a particle entering the meniscus is entrained if and 
only if its radius is smaller than the meniscus at the stagnation point 
\cite[][]{sauret2019capillary}, so that a capture probability function can be approximated as
\begin{equation}
    p_{c}^{*}(x, d)=H\left(h^{*}-\frac{d}{2}-x\right),
\end{equation}
where $H(x)$ is the Heaviside step function. Note that other ansatz for $p_{c}^{*}(x, d)$ could be used to describe the entrainment of more complex particles (emulsion droplets, deformable capsules, or anisotropic particles). We also assume the following expression for the particle-surface pair correlation function:
\begin{equation}
g_{s}^{*}(x, d) \approx \bar{g}_{s}^{*}(d) H(x-d/2),
\end{equation}
where the Heaviside function takes into account excluded volume near the surface ($x<d/2$), and the constant meniscus surface correlation $\bar{g}_{s}^{*}(d)$ reflects long-range particle-surface forces or dynamical effects, such as boundary layer depletion, that rescales the particle density arriving within the meniscus region relative to the well-mixed bulk fluid. In this approximation, the volume fraction of particles of diameter $d$ entering the meniscus region (outside the excluded volume near the wall) is $\phi^{*}(d)=\bar{g}_{s}^{*}(d) \phi(d)$, which could be considered similar to the bulk volume fraction for a well-mixed suspension with $\bar{g}_{s}^{*}(d) \simeq 1$. Besides neglecting variations in particle mass transfer to the meniscus region, the following analysis also neglects interactions between particles that may lead to cooperative entrainment phenomena. For instance, a large particle can briefly deform the interface at the meniscus so that nearby smaller particles are more easily entrained. In particular, clusters of particles have been shown to be able to be collectively entrained at small film thickness \cite[][]{colosqui2013hydrodynamically,sauret2019capillary} and could, in principle, be accounted for through this function. In the following, we consider its simplest expression.

With these assumptions, Equation (\ref{eq:model_2}) reduces to the integral over part of the velocity profile in the meniscus,

\begin{equation} \label{eq:Qc}
Q_{c}(d) = \phi^{*}(d) \,P \int_{d/2}^{h^{*}-d/2} \,u_{z}^{*}(x) \,{\rm d} x
\end{equation}

We further assume an approximately parabolic velocity profile vanishing at the stagnation point,

\begin{equation} \label{eq:flow_profile}
u_{z}^{*}(x) = U\left(1-\frac{x}{h^{*}}\right)^2.
\end{equation}

This expression for the velocity field ensures the mass conservation for the case of a pure liquid going into the film, $\int_{0}^{h^{*}}\, u_{z}^{*}(x) \,{\rm d} x=U \,h$. We substitute the flow profile given by equation (\ref{eq:flow_profile}) in equation (\ref{eq:Qc}) and perform the integral to obtain the probability distribution of particles in the coating $\phi_{\rm c}(d)$, defined as the volume fraction of particles of diameter $d$:
\begin{equation} \label{eq:filtration}
\phi_{\rm c}(d)=\frac{Q_{\rm c}(d)}{Q_{\rm f}}=\phi(d)\, f(\tilde{d}) \,H(1- \tilde{d}),
\end{equation}
where $\phi(d)$ is the probability distribution of particles in the suspension bath, $Q_{\rm f}$ is the flow rate of liquid in the coating film, and the Heaviside function $H(1- \tilde{d})$ indicates a sharp size cutoff given by entrainment criterion $\tilde{d}={d}/{2\,h^{*}}<1$. In equation (\ref{eq:filtration}), the filtration function defining the ratio of final to initial probability distribution of particles is given by
\begin{equation}
f(\tilde{d})=(1-\tilde{d})^{3}-\tilde{d}^{3}
\end{equation}
where we have introduced a dimensionless particle radius as
\begin{equation}
\tilde{d}=\frac{d}{2\,h^{*}}=\frac{d}{5.64\, \ell_c\, {\mathrm{Ca}_\phi}^{2 / 3}}.
\end{equation}

Note that we have used here a thickness at the stagnation point calculated with the effective viscosity of the suspension. However, because of the similar size between the particles and the stagnation point, the local thickness may be modified by the volume fraction, and the size of the particles as described recently in the wetting dynamics by \cite{zhao2020spreading}. Further experiments focusing on the exact structure and local composition of the suspension at the meniscus would be needed to refine this assumption. The thickness in this region will be set by the deformation of the meniscus, the viscosity in the bath, and the ratio of particle size to the typical lengthscale. This assumption could lead to small discrepancies in quantitative estimating the number density of entrained particles. As expected, $f(0)=1$ so that all particles that are small compared to the film thickness will be entrained if they arrive at the meniscus. The number of entrained particles of diameter $d$ per unit area in the coating is thus given by
\begin{equation}
n_{p}(d)=\frac{Q_{c}(d)}{U\,P\,V_{p}(d)}=\frac{\phi_{c}(d) \,h}{V_{p}(d)}=\frac{\phi(d)\, f(\tilde{d}) \,H(1- \tilde{d})\,h}{V_{p}(d)}
\end{equation}
where $V_{p}(d)=\pi d^{3}/6$ is the volume of the spherical particle, and $h$ is the film thickness. The total entrained solid volume fraction and total solid flow rate (entrained volume per time) in the coating are expressed as
\begin{equation}
    \phi_{p}=\int_{0}^{\infty} \phi_{c}(d) \,{\rm  d} d \quad {\rm and} \quad Q_{s}=\int_{0}^{\infty} Q_{c}(d)\,{\rm  d} d=\phi_{p} Q_{f}.
\end{equation}

We apply these equations to the particular case of the bidisperse suspensions used in this study. We can express the probability distribution of small and large particles as $\phi_{\rm S}=(1-\zeta)\,\phi$ and $\phi_{\rm L}=\zeta\,\phi$, respectively. As a result, the number of small particles of radius $d_{\rm S}$ entrained per unit area is
\begin{equation}\label{eq:model_large}
n_{p}(d_{\rm S})=\frac{(1-\zeta)\,\phi\, f(\tilde{d}_{\rm S}) \,H(1- \tilde{d}_{\rm S})\,h}{V_{p}(d_{\rm S})},
\end{equation}
and for the large particles of diameter $d_{\rm L}$
\begin{equation}\label{eq:model_small}
n_{p}(d_{\rm L})=\frac{\zeta\,\phi\, f(\tilde{d}_{\rm L}) \,H(1- \tilde{d}_{\rm L})\,h}{V_{p}(d_{\rm S})}.
\end{equation}

\begin{figure}
\centering
\subfigure[]{\includegraphics[width=0.495\textwidth]{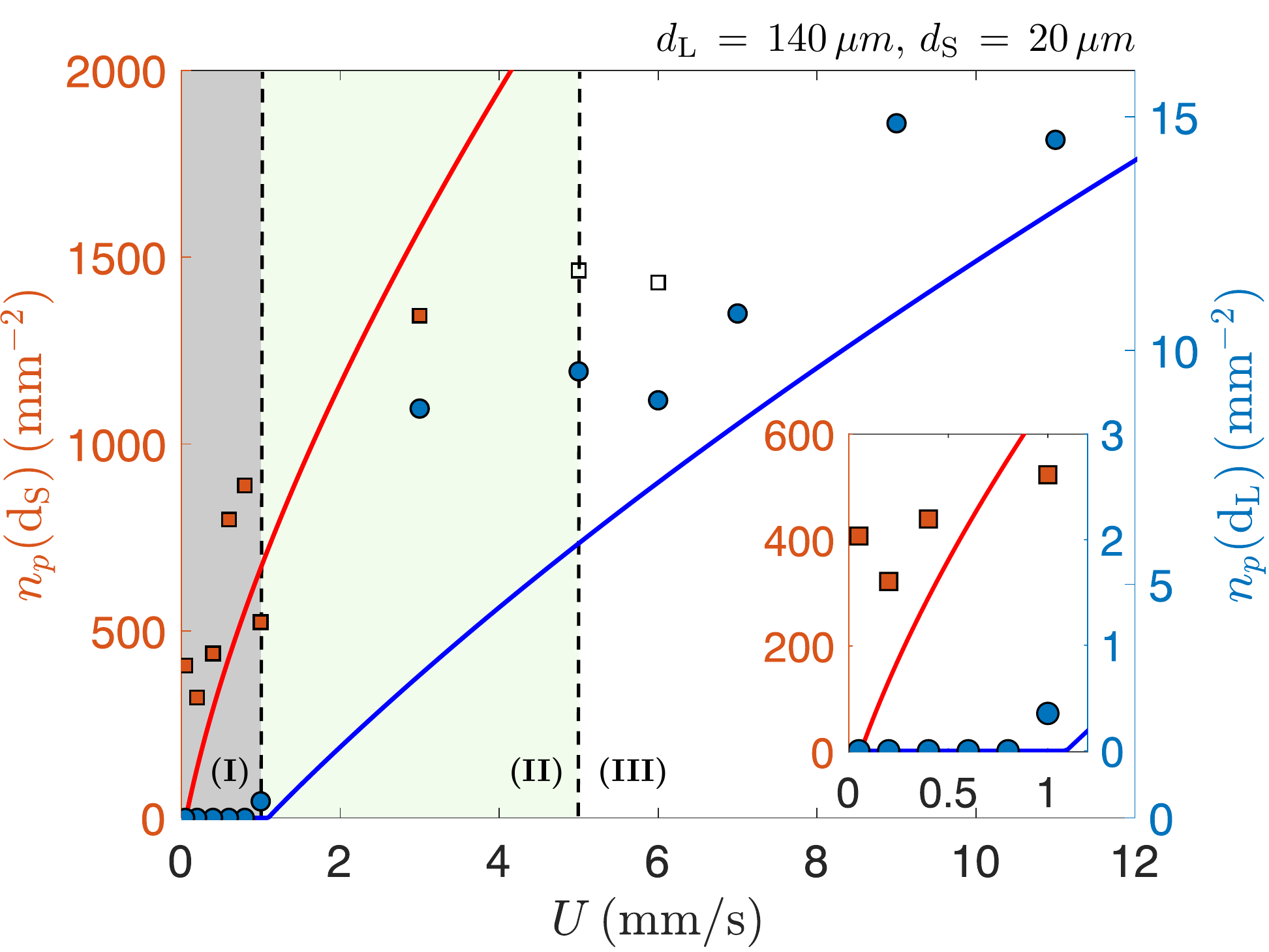}}
\subfigure[]{\includegraphics[width=0.495\textwidth]{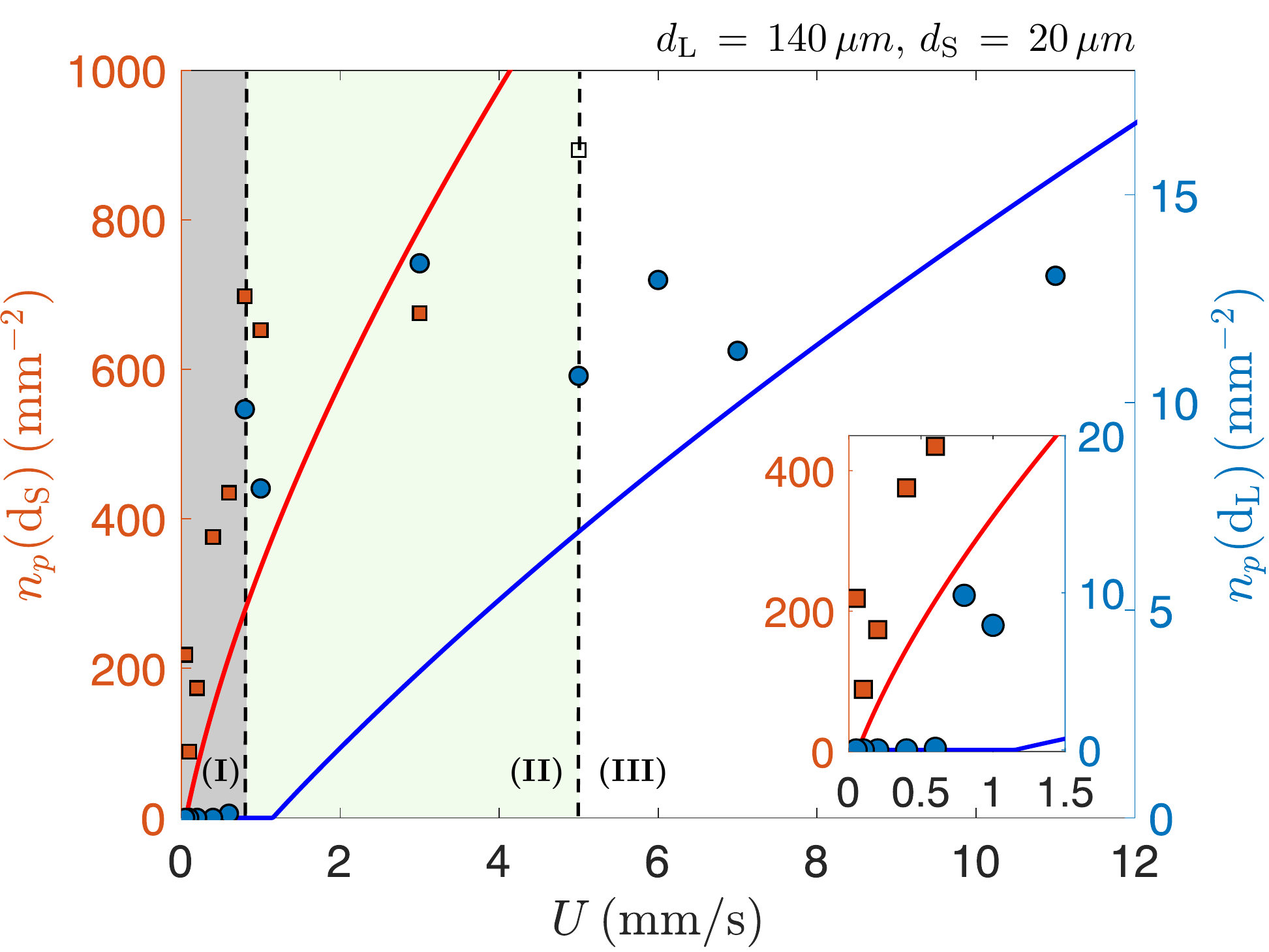}}
\subfigure[]{\includegraphics[width=0.495\textwidth]{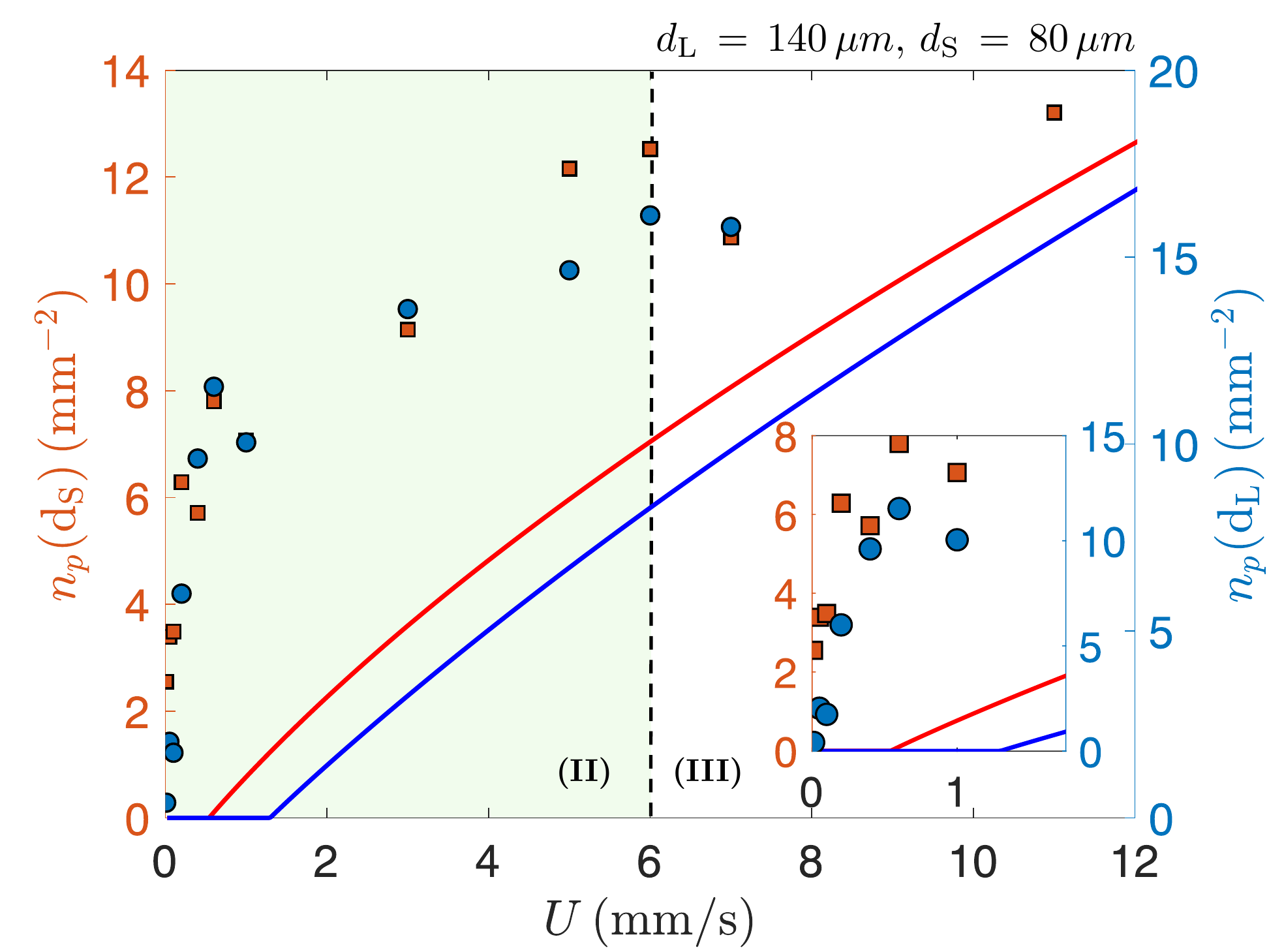}}
\subfigure[]{\includegraphics[width=0.495\textwidth]{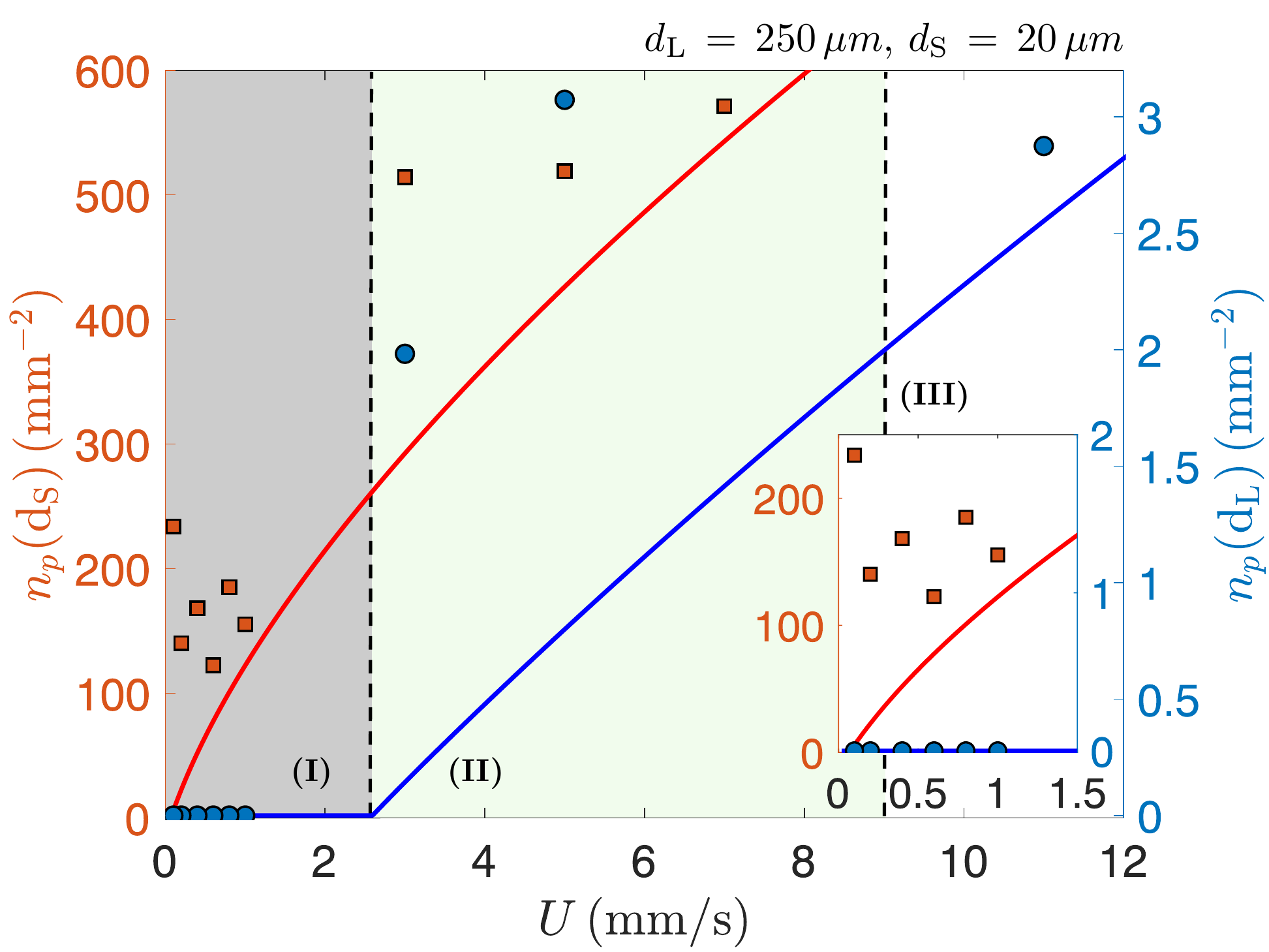}}
  \caption{Number of particles deposited on a unit area of the plate.
      The composition of the suspension is 
      (a) $\phi=0.2$, $\zeta=0.6$, $\delta=7$ 
      ($d_{\rm S}=20\,\mu{\rm m}$ and $d_{\rm L}=140\,\mu{\rm m}$); 
      (b) $\phi=0.2$, $\zeta=0.8$, $\delta=7$ 
      ($d_{\rm S}=20\,\mu{\rm m}$ and $d_{\rm L}=140\,\mu{\rm m}$); 
      (c) $\phi=0.2$, $\zeta=0.9$, $\delta=1.75$ 
      ($d_{\rm S}=80\,\mu{\rm m}$ and $d_{\rm L}=140\,\mu{\rm m}$); 
      (d) $\phi=0.2$, $\zeta=0.9$, $\delta=12.5$ 
      ($d_{\rm S}=20\,\mu{\rm m}$ and $d_{\rm L}=250\,\mu{\rm m}$).
      The blue and red symbols correspond to the experimental measurements for the large and small particles, respectively.
      The open squares represent the cases where the number of particles is underestimated because of limitations in the camera's depth-of-field and multi-layer deposition.
      The solid blue and red lines correspond to the theoretical prediction of the number of large
      [equation (\ref{eq:model_large})] and small [equation (\ref{eq:model_small})] particles per unit area on the plate, respectively.
      Insets: Zoomed-in view on low withdrawal velocity region highlighting the ability of filtering by size the particles in the suspension.
  }
  \label{fgr:Figure_9}
\end{figure}

These expressions are plotted in figures \ref{fgr:Figure_8}(a)-(d) and show a fair agreement with the experimental data. The model yields quantitative results as it gives a reasonable estimate of the number of entrained particles when varying the volume ratio of large particles $\zeta$ from $0.6$ [figure  \ref{fgr:Figure_8}(a)] to $0.8$ [figure~\ref{fgr:Figure_8}(b)]. In both cases, the main limit occurs near the threshold velocity, where the model underpredicts the number of particles entrained. Indeed, the threshold for entrainment is based on the criterion for individual particles \cite[][]{colosqui2013hydrodynamically,sauret2019capillary}, and does not account for the clustering of particles that deform the meniscus and allows particles at sufficient volume fraction to be entrained earlier. It was previously reported that the onset of the monolayer regime depends significantly on the volume fraction and so far remains empirically measured \cite[][]{palma2019dip}. Adding this component to the model presented above could lead to a better quantitative prediction of the density of entrained particles in this region. Nevertheless, the number density of particles is significantly filtered in this heterogeneous regime.

We have also performed similar measurements with smaller size differences  ($\delta=1.75$ in figure \ref{fgr:Figure_8}(c)). The experiments show that the velocity range in which particles could be separated by size is significantly reduced. In addition, figure \ref{fgr:Figure_8}(c) shows that for similar particle sizes, an heterogeneous regime for both particle sizes is quickly reached and the prediction does not capture well the number of particles entrained per unit area. On the other hand, for large size difference ($\delta=12.5$ in figure \ref{fgr:Figure_8}(d)), there is a clear range below $U=2\,{\rm mm/s}$ in which the film is free of large particles while entraining the small particles.

These results illustrate that during the coating of a plate with a polydisperse suspension, the composition of the coating film may be very different compared to the composition of the bath. In the heterogeneous regime, the resulting coating will contain more small particles and fewer large particles than the original composition of the suspension, possibly compromising the quality of the coating.

\subsection{Using dip-coating as a filtration method}

The separation of small particles in the micron-size range (up to 1000 microns) from a liquid dispersion is a source of challenge. When decreasing the particle size and increasing the batch volume, most filtering methods are neither very efficient nor suitable for a large throughput and/or are not highly selective. For example, the use of mechanical filters of specific pore size can quickly lead to clogging of the filter pores that slow down the filtering process \citep[][]{urfer1997biological,wyss2006mechanism,dressaire2017clogging,sauret2018growth}. Centrifugation is also a standard filtering method but cannot separate the particles with a high selectivity \cite[][]{svarovsky2000solid}. Besides, centrifugation relies on the difference of density between the particles, and if the densities are comparable, the process loses in efficiency \cite[][]{ninfa2009fundamental}. However, a filtration method through a dip-coating process, as demonstrated in the insets of figure \ref{fgr:Figure_9}(a)-(d), depends on whether or not a particle enters in the coating film, which is mainly governed by the diameters of the particles and could therefore be used to sort particles by size, regardless of the particle volume fraction as reported previously \cite[][]{dincau2019capillary}. The main limitation to this method is that it is desirable to have particles with a significant size difference.

The results demonstrated here with solid particles and a bimodal distribution could be extended to particles with a polydisperse size distribution and different material types. In addition, whereas the size of the entrained particles is limited here by the thickness of the liquid film, which is directly correlated to the capillary length, this filtration method could also be used with, and fibers \cite[][]{dincau2020entrainment}. In this case, the thickness of the film, and thus the thickness at the stagnation point, is directly proportional to the radius of the fiber and therefore allows a larger range of size of particles that could be filtered. 

In summary, the present methods offer various potential applications: in medicine where it may be used to separate blood plasma components and cells (in a range of size from $5$ to $50\,\mu{\rm m}$) and for grains and powders, such as ceramic abrasives, where the standard methods of sedimentation or centrifugation are relatively slow, inaccurate and costly. Here, the possibility of scaling up the capillary filtering mechanisms with arrays of wires could open the opportunity for high throughput and good efficiency.

\section{Conclusions}

In this paper, we have investigated the dip-coating of a plate withdrawn from a bath containing a suspension of particles with a bimodal size distribution. Previous studies have reported that different regimes are observed for monodisperse suspensions \cite[][]{sauret2019capillary,palma2019dip}: no entrainment at low velocities; a heterogeneous regime with a monolayer of particles at intermediate velocity; and an effective viscosity regime when the film is thicker than the particle diameter. For bidisperse suspensions, the difference in diameter of the particles dispersed in the suspensions introduced a new complexity as additional length scales need to be compared to the film thickness. 

The behavior observed for bidisperse suspensions is summarized in figure \ref{fgr:Figure_1}. At low velocity and moderate volume fraction, barely any particles are entrained. Increasing the withdrawal velocity leads to a peculiar behavior: initially, only the small particles are entrained on the plate, and the coating exhibits a heterogeneous regime with a monolayer of small particles. Increasing the withdrawal velocity further leads to a second velocity threshold where the large particles start to be entrained. The number of entrained large particles gradually increases with the withdrawal velocity. Finally, at large enough withdrawal velocities, the composition of the suspension in the coating film is mostly similar to the composition of the suspension in the bath. A model that accounts for the probability of entraining particles based on their size has been developed. The model qualitatively reproduces the experimental measurements, although the heterogeneous regime where particles are entrained collectively is not captured in this model. 

Our experiments have also revealed that the size of the largest particles in the suspension controls the onset of the effective viscosity regime. When the thickness of the coating film becomes larger than the diameter of the large particles, $h \geq d_{\rm L}$, the thickness can be predicted at first order by the Landau-Levich-Derjaguin (LLD) law by considering the effective viscosity of the bidisperse suspension. The presence of the different particle sizes, however, lowers the viscosity of the suspension for a given volume fraction $\phi$, and the evolution of the viscosity is well-predicted by a model that considers the polydispersity and its influence on the maximum packing fraction. We should emphasize that both in the monodisperse and bidisperse regimes, the volume fraction of particles in the coating film always exhibits a slight decrease compared to the volume fraction in the liquid bath, likely due to a self-filtering mechanism that deserves further investigations.

Our experiments revealed that the dip coating process is an efficient tool to control the coating of plates passively by tuning the thickness of the coating film, as illustrated in this paper. This approach also provides a new method to sort particles by size in polydisperse suspensions.

\section*{Author Contributions}
DHJ: Methodology, Investigation, Formal analysis, Data curation, Visualization, Writing – original draft; MKHL: Investigation, Writing – review \& editing; VT: Methodology, Writing – review \& editing; MZB: Conceptualization, Methodology, Writing – review \& editing; AS: Conceptualization, Methodology, Supervision, Writing – review \& editing, Funding acquisition, Project administration.

\section*{Conflicts of interest}
There are no conflicts to declare.

\section*{Acknowledgements}
This material is based upon work supported by the National Science Foundation under NSF CAREER Program Award CBET Grant No. 1944844. The authors thank collaborators at Saint-Gobain Research North America, and in particular Yinggang Tian, J. Alex Lee, and Chuanping Li.

\bigskip

\appendix

\noindent \textbf{Appendix A. Physical properties of the polystyrene particles}

\smallskip

\begin{figure}
 \begin{center}
 \includegraphics[width=0.65\textwidth]{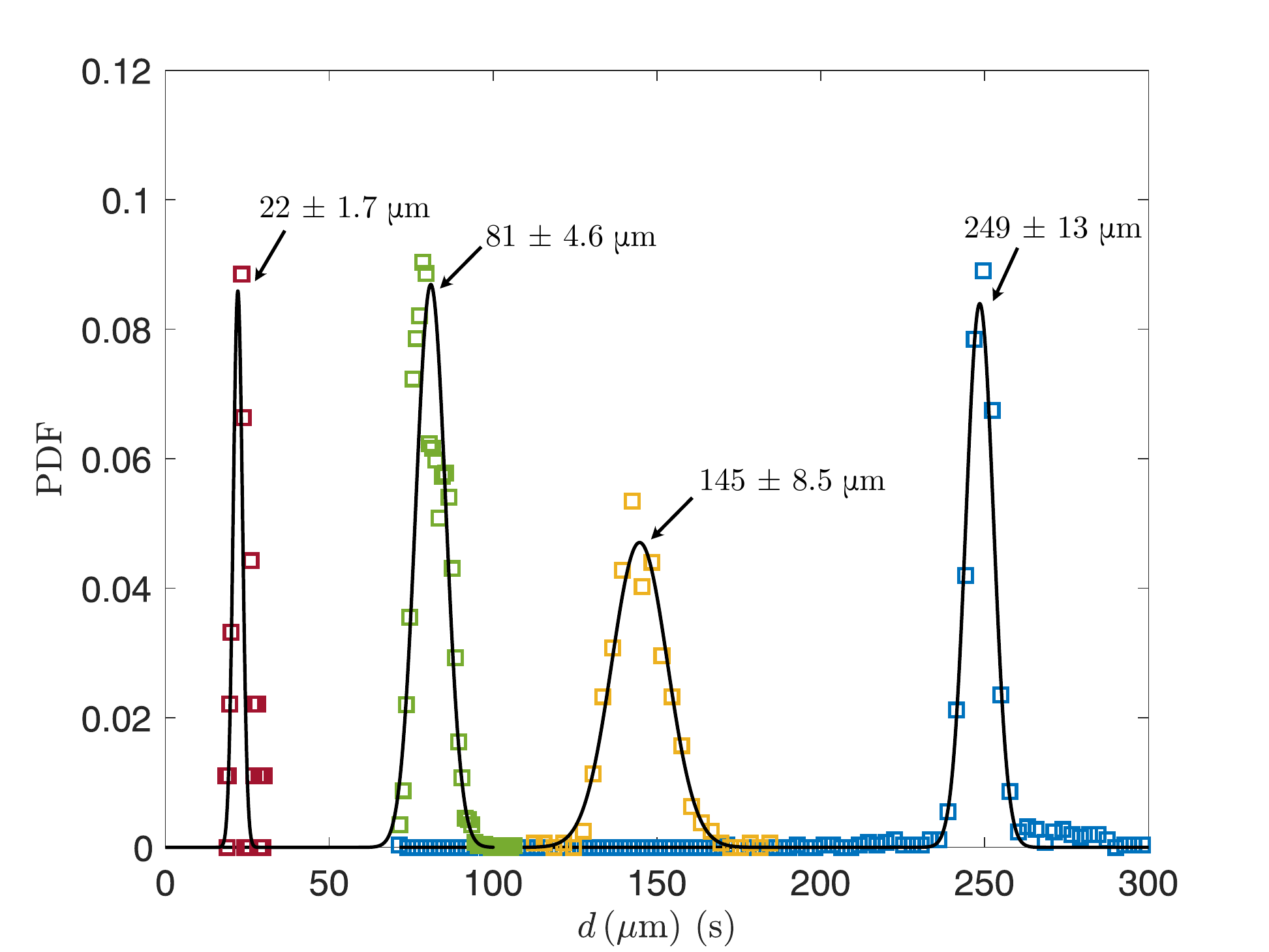}%
 \caption{Probability density function (PDF) of diameter of the polystyrene particles used in this study and corresponding Gaussian fits.\label{figure:PDF}}
  \end{center}
 \end{figure}

\noindent The particulate suspensions used in this study consist of spherical polystyrene particles (Dynoseeds TS, Microbeads) dispersed in silicone oil (AP100, Sigma Aldrich.) Particles with four different sizes have been used. We have measured the size distribution of each batch of the particles. Pictures of a large number of particles are taken and processed through ImageJ to obtain the projected area $A$ of each particle. The diameter of the particles $d$ was then obtained from the project area, $d=2\sqrt{A/\pi}$, considering that the circularity of the particles is close to 1. The measured size and standard deviation of each batch of particles are reported in table \ref{tab:properties}. The probability density function of the size distribution is plotted and fitted with a Gaussian distribution curve in Figure \ref{figure:PDF}.

The density of the particles was measured by mixing a batch of particles into salt waters with known densities. The density of each batch of particles is also reported in Table \ref{tab:properties}.

\vspace{5mm}

\begin{table}
\centering
\begin{tabular}{ccccc}
                                          & TS 20 & TS 80 & TS 140 & TS 250 \\
Mean diameter $d\,(\mu{\rm m})$           & 22    & 81    & 145    & 249    \\
Standard deviation $\sigma\,(\mu{\rm m})$ & 1.7   & 4.6   & 8.5    & 13     \\
Density $\rho$ (${\rm kg.m^{-3}}$)        & 1.046 & 1.048 & 1.060  & 1.062 
\end{tabular}
\caption{Mean diameter and density of the polystyrene particles used in this study}
\label{tab:properties}
\end{table}

\bigskip
\noindent \textbf{Declaration of Interests.} The authors report no conflict of interest
\bibliographystyle{jfm}

\bibliography{jfm-instructions}

\end{document}